\documentclass[12pt]{article}

\newcommand{\newsection}[1]{\section{#1}\setcounter{equation}{0}} 
\newcommand{\beq}{\begin{eqnarray}}
\newcommand{\eeq}{\end{eqnarray}}
\newcommand{\beqnn}{\begin{eqnarray*}}
\newcommand{\eeqnn}{\end{eqnarray*}}
\newtheorem{theorem}{Theorem}
\newtheorem{lemma}{Lemma}

\newcommand{\proof}{\paragraph*{{\it Proof.\/}}}

\newcommand{\qed}{\fbox{\phantom{-}}\bigskip}
\newcommand{\rd}{\partial}

\newcommand{\Tr}{\mathop{\mathrm{Tr}}}
\newcommand{\diag}{\mathop{\mathrm{diag}}}
\newcommand{\tp}[1]{\:{}^{\mathrm{t}}#1}
\newcommand{\Ker}{\mathop{\mathrm{Ker}}}
\newcommand{\Coker}{\mathop{\mathrm{Coker}}}
\newcommand{\Res}{\mathop{\mathrm{Res}}}
\newcommand{\CC}{\mathbf{C}}
\newcommand{\PP}{\mathbf{P}}

\newcommand{\ZZ}{\mathbf{Z}}
\newcommand{\bfalpha}{\mbox{\boldmath$\alpha$}}
\newcommand{\bfbeta}{\mbox{\boldmath$\beta$}}
\begin{document}

%%%%%%%%%%%%%%%%
%% title page %% 
%%%%%%%%%%%%%%%%

\title{Tyurin parameters and elliptic analogue of \\
nonlinear Schr\"odinger hierarchy}
\author{Kanehisa Takasaki\\
{\normalsize Graduate School of Human and Environmental Studies, 
Kyoto University}\\
{\normalsize Yoshida, Sakyo, Kyoto 606-8501, Japan}\\
{\normalsize E-mail: takasaki@math.h.kyoto-u.ac.jp}}
\date{}
\maketitle

\begin{abstract}
Two ``elliptic analogues'' of the nonlinear Schr\"odinger hiererchy 
are constructed, and their status in the Grassmannian perspective 
of soliton equations is elucidated.  In addition to the usual fields 
$u,v$, these elliptic analogues have new dynamical variables called 
``Tyurin parameters,'' which are connected with a family of vector 
bundles over the elliptic curve in consideration.  The zero-curvature 
equations of these systems are formulated by a sequence of $2 \times 2$ 
matrices $A_n(z)$, $n = 1,2,\ldots$, of elliptic functions.  In addition 
to a fixed pole at $z = 0$, these matrices have several extra poles.  
Tyurin parameters consist of the coordinates of those poles and some 
additional parameters that describe the structure of $A_n(z)$'s.  
Two distinct solutions of the auxiliary linear equations are constructed, 
and shown to form a Riemann-Hilbert pair with degeneration points.  
The Riemann-Hilbert pair is used to define a mapping to an infinite 
dimensional Grassmann variety.  The elliptic analogues of the nonlinear 
Schr\"odinger hierarchy are thereby mapped to a simple dynamical system 
on a special subset of the Grassmann variety.  
\end{abstract}

\begin{flushleft}
Running head: Tyurin parameters and elliptic analogue\\
arXiv:nlin.SI/0307030
\end{flushleft}
\newpage

%%%%%%%%%%%%%%%
%% main text %% 
%%%%%%%%%%%%%%%

\newsection{Introduction}

Many integrable systems are expressed 
in the form of a Lax equation 
$\rd_t A(\lambda) = [B(\lambda), A(\lambda)]$ 
or a zero-curvature equation 
$[\rd_x - A(\lambda), \; \rd_t - B(\lambda)] = 0$, 
where $A(\lambda)$ and $B(\lambda)$ are 
matrices of rational functions of the spectral 
parameter $\lambda$.  In other words, these 
Lax or zero-curvature equations are defined 
on the Riemann sphere.  Some integrable systems, 
such as the elliptic Calogero-Moser system and 
the Landau-Lifshitz equation, have a Lax or 
zero-curvature representation defined on a torus, 
i.e., a complex elliptic curve.  One will naturally 
expect to find a generalization to a curve of 
higher genus.  Unfortunately, it is well known 
that such a naive attempt will be confronted with 
a serious difficulty that stems from the Riemann-Roch 
theorem \cite{bib:ZM82}.  

Recently, Krichever presented a general scheme for 
constructing a Lax or zero-curvature equation on 
an algebraic curve $\Gamma$ of arbitrary genus \cite{bib:Kr02}. 
A central idea is to allow the matrices $A,B$ to have 
extra ``movable'' poles $\gamma_1,\ldots,\gamma_{rg} 
\in \Gamma$, where $r$ is the size of the matrices.  
Moreover, the matrices $A,B$ at these poles are 
assumed to have a special structure.  A set of 
additional parameters are introduced to parametrize 
this special structure.  The coordinates of poles 
and these parameters are called ``Tyurin parameters.''  
This notion originates in algebraic geometry of 
holomorphic vector bundles over algebraic curves 
\cite{bib:Ty67}, and was applied by Krichever and Novikov 
in 1970's to the study of commutative rings of 
differential operators \cite{bib:Kr78,bib:KN78,bib:KN80}. 
The aforementioned difficulty can be resolved by 
adding Tyurin parameters as new dynamical variables.  

Once applied to zero-curvature equations, Krichever's method 
yileds a large class of $1 + 1$ dimensional integrable PDE's.  
These equations are to be called ``soliton equations'' 
associated with an algebraic curve (though it is not known 
whether these equations do have a soliton or soliton-like 
solution).  For instance, Krichever illustrates his construction 
for the case of a ``field analogue'' of the elliptic 
Calogero-Moser system.  This raises a natural question: 
What is the status of these new equations in 
the Grassmannian perspective of soliton equations due to 
Sato \cite{bib:SS82} and Segal and Wilson \cite{bib:SW85}?  

We address this problem in a simplified setting, namely, 
zero-curvature equations of $2 \times 2$ matrices defined 
on an elliptic curve.  This system is an analogue of 
the usual nonlinear Schr\"odinger hierarchy.  
More precisely, we construct two distinct versions of this 
``elliptic analogue,'' one being based on Krichever's idea, 
and the other inspired by the work of Enriquez and Rubtsov 
\cite{bib:ER99}.  Whereas Krichever's construction 
requires all Tyurin parameters to be dynamical variables, 
Enriquez and Rubtsov keeps the position of poles constant 
and use the other parameters as dynamical variables.  
In this respect, the elliptic analogue \`a la Enriquez 
and Rubtsov's is much closer to usual soliton equations.  

Our strategy is, firstly, to derive a kind of Riemann-Hilbert 
problem for these systems, and secondly, to translate it to 
the language of an infinite dimensional Grassmann variety.  
This is indeed the procedure that has been used in the literature 
for many soliton equations and some higher dimensional systems; 
see, e.g., the book of Mason and Woodhouse \cite{bib:MW96}.  
The usual Riemann-Hilbert problem, however, does not work 
literally in the present situation.  Whereas the usual 
Riemann-Hilbert problem is based on {\it triviality} of 
a holomorphic vector bundle over the Riemann sphere, 
the systems formulated by Tyurin parameters are obviously 
related to a {\it nontrivial} holomorphic vector bundle 
over an algebraic curve of positive genus.  An answer to 
this puzzle can be found in the work of Krichever and 
Novikov \cite{bib:Kr78,bib:KN78,bib:KN80} cited above.  
They consider a Riemann-Hilbert problem with 
{\it degeneration points\/}; Tyurin parameters are 
nothing but the geometric data of those points.  
The next task is, therefore, to connect this kind of 
Riemann-Hilbert problems with an infinite dimensional 
Grassmann variety.  Fortunately, a related issue has been 
investigated by Previato and Wilson \cite{bib:PW89}.  
They demonstrate therein a Grassmannian version of 
the ``dressing method'' --- a classical technique in 
soliton theory --- to solve a Riemann-Hilbert problem 
of the same type.  Moreover, their paper shows what should 
be the ``vacuum'' (to be ``dressed'') that corresponds to 
a holomorphic vector bundle in the Tyurin parametrization.  
Our goal is to develop a similar machinery for the present setting. 

This paper is organized as follows.  
Section 2 is a brief review of the usual nonlinear 
Schr\"odinger hierarchy.  This will serves as a prototype 
of the subsequent construction.  Section 3 is devoted to 
the construction of the first version, \`a la Krichever, 
of the elliptic analogues.  A technical clue is a generating 
function $U(z)$, which has been used for the usual nonlinear 
Schr\"odinger hierarchy  as well.  This enables one to 
formulate the generators of time evolutions systematically.  
Section 4 deals with an auxiliary linear system of 
the hierarchy and a pair of solutions thereof.  This pair 
of solutions turns out to satisfy a Riemann-Hilbert problem 
with degeneration points on the elliptic curve.  
Section 5 presents main results of this paper, namely, 
a Grassmannian perspective of the elliptic analogue of 
the nonlinear Schr\"odinger hierarchy.   An infinite dimensional 
Grassmann variety $\mathrm{Gr}$, a special basepoint 
(``vacuum'') $W_0 \in \mathrm{Gr}$ and the set $\mathcal{M} 
\subset \mathrm{Gr}$ of ``dressed vacua'' are introduced.  
The Riemann-Hilbert pair determines a point of $\mathcal{M}$, 
whose motion turns out to obey a simple exponential law.  
The elliptic nonlinear Schr\"odinger hierarchy is thus 
mapped to a dynamical system on $\mathcal{M}$.  
In Section 6, the same story is repeated for 
the elliptic analogue \`a la Enriquez and Rubtsov.  
Our conclusion is shown in Section 7.

\newsection{Nonlinear Schr\"odinger hierarchy}

As a prototype of the elliptic analogue, we here review 
a standard construction of the nonlinear Schr\"odinger hierarchy.  
Generalities and backgrounds of this kind of construction 
of soliton equations can be found in Frenkel's lectures 
\cite{bib:Fr98}.

\subsection{$A$-matrix}

The construction starts from the $A$-matrix 
\beq
  A(\lambda) = 
  \left(\begin{array}{cc} 
  \lambda & u \\
  v & - \lambda 
  \end{array}\right) 
\eeq
with a rational spectral parameter 
$\lambda \in \PP^1$; $u$ and $v$ are fields on 
the $x$ space.  In view of the homogeneous 
grading of an underlying loop algebra, 
it is natural to express this matrix as 
\beq
  A(\lambda) = J\lambda + A^{(1)} 
\eeq
where 
\beqnn
  J = \left(\begin{array}{cc} 
      1 & 0 \\
      0 & -1 
      \end{array}\right), \quad 
  A^{(1)} = 
      \left(\begin{array}{cc} 
      0 & u \\
      v & 0 
      \end{array}\right). 
\eeqnn

\subsection{Generating functions} 

A clue of the construction of the hierarchy is 
a Laurent series 
\beqnn
  U(\lambda) = \sum_{n=0}^\infty U_n\lambda^{-n}, \quad 
  U_0 = J, 
\eeqnn
that satisfies the differential equation 
\beq
  [\partial_x - A(\lambda),\; U(\lambda)] = 0. 
  \label{eq:nls-Lax-x-U}
\eeq
Although this equation itself does not determine 
$U(\lambda)$ uniquely, there is a {\it good} 
or {\it canonical} solution that takes the form 
\beq
  U(\lambda) = \phi(\lambda)J\phi(\lambda)^{-1}, 
  \label{eq:nls-U-Phi}
\eeq
where $\phi(\lambda)$ is a Laurent series of 
the form 
\beqnn
  \phi(\lambda) = I + \sum_{n=1}^\infty \phi_n\lambda^{-n}
\eeqnn
and satisfies the differential equations 
\beq
  \rd_x\phi(\lambda) 
  = A(\lambda)\phi(\lambda) - \phi(\lambda)J\lambda. 
  \label{eq:nls-Sato-x-Phi}
\eeq
A solution of (\ref{eq:nls-Lax-x-U}) of this form is 
indeed ``good'' or ``canonical'' in the sense that 
the coefficients $U_n$ can be calculated from $A(\lambda)$ 
by a purely algebraic procedure (namely, without actually 
solving differential equations) as follows. 
Expanded in powers of $\lambda$, (\ref{eq:nls-Lax-x-U}) 
becomes a system of differential equations 
\beqnn
  \rd_x U_n = JU_{n+1} - U_{n+1}J + [A^{(1)},U_n] 
\eeqnn
for the coefficients $U_n$.  On the other hand, 
if $U(\lambda)$ is written as (\ref{eq:nls-U-Phi}), 
the algebraic constraint 
\beq
  U(\lambda)^2 = I 
\eeq
is automatically satisfied.  This yields 
the algebraic relations 
\beqnn
  0 = JU_{n+1} + U_{n+1}J + \sum_{m=1}^n U_m U_{n+1-m} 
\eeqnn
of $U_n$'s.  One can use these relations to eliminate 
the term $U_{n+1}J$ on the right hand sides of 
the foregoing differential equations. The outcome 
are the recurrence relations 
\beq
  2JU_{n+1} = \rd_x U_n - [A^{(1)},U_n] 
              - \sum_{m=1}^n U_m U_{n+1-m} 
\eeq
that determine $U_n$'s successively as 
\beqnn
  U_1 = \left(\begin{array}{ll} 
        0 & u \\
        v & 0 
        \end{array}\right), \quad 
  U_2 = \left(\begin{array}{ll} 
        - \frac{1}{2}uv & \frac{1}{2}u_x \\
        - \frac{1}{2}v_x & \frac{1}{2}uv 
        \end{array}\right), 
\eeqnn
etc.  Note that the matrix elements of all $U_n$'s 
thus turn out to be ``local'' quantities, namely, 
polynomials of $x$-derivatives of $u$ and $v$.  

The coefficients of $\phi(\lambda)$ are ``nonlocal.''  
To construct $\phi(\lambda)$ from $A(\lambda)$, 
one expands (\ref{eq:nls-Sato-x-Phi}) to 
the differential equations 
\beq
  \rd_x \phi_n = [J, \phi_{n+1}] + A^{(1)}\phi_n 
\eeq
for the coefficients and solves them step by step. 
Actually, this is not so straightforward;   
one has to split $\phi_n$ into the diagonal and 
off-diagonal parts, 
\beqnn
  \phi_n = (\phi_n)_{\mathrm{diag}} 
         + (\phi_n)_{\mathrm{off-diag}}, 
\eeqnn
and consider them separately.  The differential equation 
for the coefficients of $\phi(\lambda)$ are thereby 
decomposed to the two equations 
\beq
  \rd_x(\phi_n)_{\mathrm{diag}} 
  = (A^{(1)}\phi_n)_{\mathrm{diag}} 
  \label{eq:phi-diag}
\eeq
and 
\beq
  \rd_x(\phi_{n-1})_{\mathrm{off-diag}} 
  = [J,\; (\phi_n)_{\mathrm{off-diag}}] 
    + (A^{(1)}\phi_{n-1})_{\mathrm{off-diag}}. 
  \label{eq:phi-offdiag}
\eeq
(For convenience, the index $n$ in the second equation 
has been shifted.)  The first equation determines 
$(\phi_n)_{\mathrm{diag}}$, up to integration constants, 
if $\phi_1,\ldots,\phi_{n-1}$ and $(\phi_n)_{\mathrm{off-diag}}$ 
are given.  The second equation is rather an algebraic equation 
that determines $(\phi_n)_{\mathrm{off-diag}}$ from 
$\phi_1,\cdots,\phi_{n-1}$.  To construct a solution, 
therefore, one has to use these equations in a cyclic way: 
\begin{enumerate}
\item Solve (\ref{eq:phi-offdiag}) for $(\phi_n)_{\mathrm{off-diag}}$. 
\item Solve (\ref{eq:phi-diag}) for $(\phi_n)_{\mathrm{diag}}$. 
\item Increase $n$ by $1$ and return to step 1. 
\end{enumerate}
The first step of this cycle is to construct 
$(\phi_1)_{\mathrm{off-diag}}$ as a solution of 
(\ref{eq:phi-offdiag}) ($n = 1$); note that 
the only data necessary here is $\phi_0 = I$.  
Starting with this step, one can proceed as 
$(\phi_1)_{\mathrm{off-diag}}$ $\to$ 
$(\phi_1)_{\mathrm{diag}}$ $\to$ 
$(\phi_2)_{\mathrm{off-diag}}$ $\to$ 
$(\phi_2)_{\mathrm{diag}}$ $\to$ $\cdots$.  
Changing integration constants in the solution 
of  (\ref{eq:phi-diag}) amounts to the right action 
$\phi(\lambda) \to \phi(\lambda)C(\lambda)$ 
by a diagonal matrix $C(\lambda) 
= \diag(c_1(\lambda),c_2(\lambda))$ of 
Laurent series with constant coefficients.

\subsection{Construction of hierarchy}

Having constructed the generating function $U(\lambda)$, 
one can formulate the hierarchy as the system of 
the Lax equations 
\beq
  [\rd_{t_n} - A_n(\lambda),\; U(\lambda)] = 0, 
\eeq
where $A_n(\lambda)$ denotes the ``polynomial part'' 
of $U(\lambda)\lambda^n$: 
\beq
  A_n(\lambda) 
  = U_0\lambda^n + U_1\lambda^{n-1} + \cdots + U_n. 
\eeq
Since $U_1 = A^{(1)}$, $A_1(\lambda)$ coincides 
with $A(\lambda)$, so that $x$ can be identified 
with the first time variable $t_1$.  As we shall show 
later in a more complicated situation, one can derive 
the zero-curvature equations 
\beq
  [\rd_{t_m} - A_m(\lambda),\; \rd_{t_n} - A_n(\lambda)] = 0 
\eeq
from these Lax equations of $U(\lambda)$.  Actually, 
another set of zero-curvature equations, i.e.,  
\beq
  [\rd_{t_m} - A_m^{-}(\lambda),\; \rd_{t_n} - A_n^{-}(\lambda)] = 0, 
\eeq
can be derived for the Laurent ``tail'' 
\beq
  A_n^{-}(\lambda) 
  = A_n(\lambda) - U(\lambda)\lambda^n 
  = - U_{n+1}\lambda^{-1} - U_{n+2}\lambda^{-2} - \cdots 
\eeq
as well.  These ``dual'' zero-curvature equations 
are the Frobenius integrability condition of 
the linear system 
\beq
  \rd_{t_n}\phi(\lambda) = A_n^{-}(\lambda)\phi(\lambda), 
\eeq
which thereby determine the time evolutions of $\phi(\lambda)$.  
This linear system turns out to be equivalent to 
the usual auxiliary linear system 
\beq
  \rd_{t_n}\psi(\lambda) = A_n(\lambda)\psi(\lambda) 
\eeq
upon identifying 
\beqnn
  \psi(\lambda) 
  = \phi(\lambda)\exp\Bigl(\sum_{n=1}^\infty t_nJ\lambda^n\Bigr) 
  \quad (t_1 = x).  
\eeqnn

\newsection{Construction of elliptic analogue \`a la Krichever} 

Let $\Gamma$ be a nonsingular elliptic curve realized 
as the torus $\CC/(2\omega_1\ZZ + 2\omega_3\ZZ)$, 
and $z$ the complex coordinate of $\CC$, which is 
also understood as a local coordinate of $\Gamma$.  
The polynomial matrices $A(\lambda),A_n(\lambda)$ in 
the nonlinear Schr\"odinger hierarchy are replaced by 
matrices $A(z),A_n(z)$ of meromorphic functions on $\Gamma$.  
They have a fixed pole at $z = 0$ (which amounts to 
$\lambda = \infty$ in the nonlinear Schr\"odinger hierarchy) 
and two ``movable'' poles at $z = \gamma_1,\gamma_2$, 
$\gamma_1 \not= \gamma_2$.

\subsection{$A$-Matrix on elliptic curve}

The role of the $A$-matrix in the usual 
nonlinear Schr\"odinger hierarchy is now played 
by a $2 \times 2$ matrix $A(z)$ ($z \in \Gamma$) 
of meromorphic functions on $\Gamma$ with the following 
properties: 
\begin{enumerate}
\item $A(z)$ has poles at $z = 0,\gamma_1,\gamma_2$ and 
is holomorphic at other points.  
\item As $z \to 0$, 
\beq
   A(z) = \left(\begin{array}{cc} 
          z^{-1} & u \\
          v & - z^{-1} 
          \end{array}\right) 
        + O(z). 
\eeq
\item As $z \to \gamma_s$, $s = 1,2$, 
\beq
  A(z) = \frac{\bfbeta_s\tp{\bfalpha_s}}{z - \gamma_s} + O(1), 
\eeq
where $\bfalpha_s$ and $\bfbeta_s$ are two-dimensional 
column vectors that do not depend on $z$.  $\bfalpha_s$ 
is normalized as $\bfalpha_s = \tp{(\alpha_s,1)}$.  
\end{enumerate}
$\gamma_s$ and $\alpha_s$ in this definition are 
the Tyurin parameters in the present setting.  
$u$ and $v$ are counterparts of those in 
the nonlinear Schr\"odinger hierarchy.  
All these parameters are understood to be dynamical, 
i.e., a function of $x$ (and the time variables $t_n$ 
to be introduced later).  We have thus altogether 
six dynamical variables 
$\gamma_1,\gamma_2,\alpha_1,\alpha_2,u,v$.  

\begin{lemma}
If $\alpha_1 \not= \alpha_2$, a matrix $A(z)$ of 
meromorphic functions on $\Gamma$ with these properties 
does exists.  It is unique and can be written explicitly 
in terms of the Weierstrass zeta function $\zeta(z)$ as 
\beq
  A(z) 
  = \sum_{s=1,2}\bfbeta_s\tp{\bfalpha_s}
      (\zeta(z - \gamma_s) + \zeta(\gamma_s)) 
  + \left(\begin{array}{cc} 
      \zeta(z) & u \\
      v & - \zeta(z) 
    \end{array}\right), 
\eeq
where 
\beq
  \bfbeta_1 
  = \frac{1}{\alpha_1 - \alpha_2}
    \left(\begin{array}{c}
      - 1 \\
      - \alpha_2 
    \end{array}\right), \quad 
  \bfbeta_2 
  = \frac{1}{\alpha_1 - \alpha_2} 
    \left(\begin{array}{c}
      1 \\
      \alpha_1 
    \end{array}\right).      
\eeq
\end{lemma}

\proof
The defining properties of $A(z)$ imply that 
$A(z)$ can be written as 
\beqnn
  A(z) = \sum_{s=1,2}\bfbeta_s\tp{\bfalpha_s}\zeta(z - \gamma_s) 
       + J\zeta(z) + C, 
\eeqnn
where $C$ is a constant matrix.  By the residue theorem, 
the coefficients have to satisfy the linear relation 
\beqnn
  \sum_{s=1,2}\bfbeta_s\tp{\bfalpha_s} + J = 0 
\eeqnn
that ensures that $A(z)$ is single valued on $\Gamma$.  
Solving these equations for $\bfbeta_s$ leads to 
the formula stated in the lemma.  On the other hand, 
matching with the Laurent expansion of $A(z)$ 
at $z = 0$ yields to the relation 
\beqnn
  A^{(1)} = 
  \sum_{s=1,2}\bfbeta_s\tp{\bfalpha_s}\zeta(- \gamma_s) 
  + C, 
\eeqnn
which determines $C$.  \qed

The Tyurin parameters $\gamma_s$ and $\alpha_s$ 
are required to satisfy the equations 
\beq
  \rd_x \gamma_s + \Tr \bfbeta_s\tp{\bfalpha_s} = 0, 
  \label{eq:rd-x-gamma} \\ 
  \rd_x \tp{\bfalpha_s} + \tp{\bfalpha_s}A^{(s,1)} 
  = \kappa_s\tp{\bfalpha_s}, 
  \label{eq:rd-x-alpha}
\eeq
where $A^{(s,1)}$ stands for the constant term 
of the Laurent expansion of $A(z)$ at $z = \gamma_s$, 
\beqnn
  A^{(s,1)} 
  = \lim_{z\to\gamma_s}\left( 
    A(z) - \frac{\bfbeta_s\tp{\bfalpha_s}}{z - \gamma_s}
    \right), 
\eeqnn
and $\kappa_s$ is a constant to be determined by 
the equation itself.  More explicitly, we have 
\beq
  \rd_x\gamma_1 = \frac{\alpha_1 + \alpha_2}{\alpha_1 - \alpha_2}, 
  \quad 
  \rd_x\gamma_2 = - \frac{\alpha_1 + \alpha_2}{\alpha_1 - \alpha_2}, 
  \label{eq:rd-x-gamma12}\\ 
  \rd_x\alpha_1 = - 2\alpha_1\zeta_{12} - v + \alpha_1^2u, 
  \quad 
  \rd_x\alpha_2 = 2\alpha_2\zeta_{12} - v + \alpha_2^2u, 
  \label{eq:rd-x-alpha12}
\eeq
where 
\beqnn
  \zeta_{12} 
  = \zeta(\gamma_1) - \zeta(\gamma_2) 
    - \zeta(\gamma_1 - \gamma_2), 
\eeqnn
and the constants $\kappa_s$ take the form 
\beq
  \kappa_s 
    = - \frac{2\alpha_s}{\alpha_1 - \alpha_2}\zeta_{12} + \alpha_s u. 
\eeq
As Krichever's lemma \cite[Lemma 5.2]{bib:Kr02} shows, 
these equations ensure that the auxiliary linear system 
$\rd_x\psi(z) = A(z)\psi(z)$ has a $2 \times 2$ matrix solution 
that is holomorphic at $z = \gamma_s$ and invertible except 
at these points.  One will notice from (\ref{eq:rd-x-gamma12}) 
and (\ref{eq:rd-x-alpha12}) that not all of the six dynamical 
variables $\gamma_1,\gamma_2,\alpha_1,\alpha_2,u,v$ are 
independent; for instance, one can solve (\ref{eq:rd-x-alpha12}) 
for $u$ and $v$ to eliminate $u$ and $v$ as auxiliary 
dynamical variables.  In the following, however, 
we shall treat these six variables on a equal footing.

\subsection{Generating functions} 

We now proceed to the construction of two 
generating functions 
\beqnn
  \phi(z) = I + \sum_{n=1}^\infty \phi_nz^n, \quad 
  U(z) = J + \sum_{n=1}^\infty U_nz^n.  
\eeqnn

The first generating function $\phi(z)$ is a Laurent series 
that satisfies the differential equation 
\beq
  \rd_x\phi(z) = A(z)\phi(z) - \phi(z)Jz^{-1}.
  \label{eq:Sato-x-Phi}
\eeq
Here $A(z)$ is understood to be its Laurent expansion 
\beq
  A(z) = Jz^{-1} + \sum_{n=1}^\infty A^{(n)}z^{n-1} 
\eeq
at $z = 0$; the first few coefficients of this 
expansion read 
\beqnn
  A^{(1)} &=& 
    \left(\begin{array}{ll} 
      0 & u \\
      v & 0 
    \end{array}\right), 
  \nonumber \\
  A^{(2)} &=&
    \frac{1}{\alpha_1 - \alpha_2} 
    \left(\begin{array}{ll}
      \alpha_1\wp(\gamma_1) - \alpha_2\wp(\gamma_2) & 
        \wp(\gamma_1) - \wp(\gamma_2) \\
      \alpha_1\alpha_2(\wp(\gamma_1) - \wp(\gamma_2)) & 
        \alpha_2\wp(\gamma_1) - \alpha_1\wp(\gamma_2) 
    \end{array}\right), 
\eeqnn
etc.

\begin{lemma}
A Laurent series solution $\phi(z)$ of 
(\ref{eq:Sato-x-Phi}) does exist.
\end{lemma}

\proof
Expanded in powers of $z$, (\ref{eq:Sato-x-Phi}) yields 
the differential equations 
\beqnn
  \rd_x \phi_n 
  = [J, \phi_{n+1}] + \sum_{m=1}^{n+1} A^{(m)}\phi_{n+1-m} 
\eeqnn
for the coefficients $\phi_n$.  One can decompose these 
equations into the diagonal and off-diagonal parts. 
The diagonal part becomes the equation 
\beqnn
  \rd_x(\phi_n)_{\mathrm{diag}} 
  = \sum_{m=1}^{n+1} (A^{(m)}\phi_{n+1-m})_{\mathrm{diag}}, 
\eeqnn
which determines the diagonal part $(\phi_n)_{\mathrm{diag}}$ 
of $\phi_n$ up to integration constants.  The off-diagonal part 
gives the algebraic relation 
\beqnn
  \rd_x(\phi_n)_{\mathrm{off-diag}} 
  = [J,\; (\phi_{n+1})_{\mathrm{off-diag}}] 
    + \sum_{m=1}^{n+1}(A^{(m)}\phi_{n+1-m})_{\mathrm{off-diag}}. 
\eeqnn
The off-diagonal part $(\phi_{n+1})_{\mathrm{off-diag}}$ 
of $\phi_{n+1}$ is thus determined from $\phi_1,\cdots,\phi_n$. 
\qed

The second generating function $U(z)$ can be 
obtained from $\phi(z)$ as 
\beq
  U(z) = \phi(z)J\phi(z)^{-1}, 
\eeq
which satisfies the differential equation 
\beq
  [\rd_x - A(z),\; U(z)] = 0, 
  \label{eq:Lax-x-U}
\eeq
and the algebraic constraint 
\beq
  U(z)^2 = I. 
\eeq
As we have seen in the case of 
the nonlinear Schr\"odinger hierarchy, 
this algebraic constraint singles out 
a unique Laurent series solution 
of (\ref{eq:Lax-x-U}), and 
the Laurent coefficients can be calculated 
by a set of recurrence relations.  

\begin{lemma}
The coefficients $U_n$ of $U(z)$ satisfy 
the recurrence relations 
\beq
  2JU_{n+1} 
  = \rd_x U_n - \sum_{m=1}^{n+1}[A^{(m)},U_{n+1-m}] 
    - \sum_{m=1}^n U_m U_{n+1-m}. 
\eeq
\end{lemma}

\proof 
(\ref{eq:Lax-x-U}) yields the differential equations 
\beqnn
  \rd_x U_n 
  = JU_{n+1} - U_{n+1}J + \sum_{m=1}^{n+1} [A^{(m)},U_{n+1-m}] 
\eeqnn
for the coefficients $U_n$.  The algebraic constraint 
$U(z)^2 = 1$ gives the algebraic relations 
\beqnn
  0 = JU_{n+1} + UJ_{n+1} + \sum_{m=1}^n U_m U_{n+1-m}. 
\eeqnn
Combining them, one obtains the recurrence relation. \qed 

One can thus calculate $U_n$'s successively from 
the Laurent coefficients $A^{(n)}$ of $A(z)$ as 
\beqnn
  U_1 &=& 
    \left(\begin{array}{ll} 
      0 & u \\
      v & 0 
    \end{array}\right), 
  \nonumber \\
  U_2 &=& 
    \left(\begin{array}{ll} 
      - \frac{1}{2}uv & \frac{1}{2}u_x \\
      - \frac{1}{2}v_x & \frac{1}{2}uv 
    \end{array}\right) 
  + \frac{1}{\alpha_1 - \alpha_2} 
    \left(\begin{array}{ll} 
      0 & \wp(\gamma_1) - \wp(\gamma_2) \\
      \alpha_1\alpha_2(\wp(\gamma_1) - \wp(\gamma_2)) & 0 
    \end{array}\right), 
\eeqnn
etc.   In particular, the matrix elements of all $U_n$'s 
turn out to be a polynomial of $x$-derivatives of 
$u,v,\gamma_s,\alpha_s$.

\subsection{Construction of hierarchy} 

Generators of time evolution are $2 \times 2$ 
matrices $A_n(z)$, $n = 1,2,\cdots$, of 
meromorphic functions on $\Gamma$  
with the following properties: 
\begin{enumerate}
\item $A_n(z)$ has poles at $z = 0,\gamma_1,\gamma_2$ and 
is holomorphic at other points. 
\item As $z \to 0$, 
\beq
  A_n(z) = U(z)z^{-n} + O(z). 
  \label{eq:An-at-z=0}
\eeq
\item As $z \to \gamma_s$, $s = 1,2$, 
\beq
  A_n(z) 
  = \frac{\bfbeta_{n,s}\tp{\bfalpha_s}}{z - \gamma_s} 
    + O(1), 
\eeq
where $\bfbeta_{n,s}$ is a two-dimensional column vector 
that does not depend on $z$.  
\end{enumerate}

\begin{lemma}
If $\alpha_1 \not= \alpha_2$, a matrix $A_n(z)$ 
of meromorphic functions on $\Gamma$ 
with these properties does exists.  It is unique 
and can be written explicitly as 
\beq
  A_n(z) 
  = \sum_{s=1,2}\bfbeta_{n,s}\tp{\bfalpha_s} 
     (\zeta(z - \gamma_s) + \zeta(\gamma_s)) 
  + \sum_{m=0}^{n-1}\frac{(-1)^{m}}{m!}
      \rd_z^{m}\zeta(z)U_{n-1-m} 
  + U_n. 
\eeq
The vectors $\bfbeta_{n,s}$ are determined by 
the linear equation 
\beq
  \sum_{s=1,2}\bfbeta_{n,s}\tp{\bfalpha_s} + U_{n-1} = 0 
\eeq
that ensures the single-valuedness of $A_n(z)$ 
on $\Gamma$.  
\end{lemma}

\proof
Repeat the same reasoning as the case of $A(z)$. 
\qed 

Solving the last linear equation, one can eventually 
find an explicit form of $A_n(z)$ .  For instance, 
$A_1(z)$ coincides with $A(z)$, and $A_2(z)$ takes the form 
\beq
  A_2(z) 
  = \sum_{s=1,2}\bfbeta_{2,s}\tp{\bfalpha_s}
    (\zeta(z - \gamma_s) + \zeta(\gamma_s)) 
  + J\wp(z) + U_1\zeta(z) + U_2, 
\eeq
where 
\beq
  \bfbeta_{2,1} 
  = \frac{1}{\alpha_1 - \alpha_2} 
    \left(\begin{array}{cc} 
      u\alpha_2 \\
      - v 
    \end{array}\right), 
  \quad 
  \bfbeta_{2,2} 
  = \frac{1}{\alpha_1 - \alpha_2} 
    \left(\begin{array}{cc} 
      - u\alpha_1 \\
      v 
    \end{array}\right). 
\eeq

Let us assume the genericity condition 
\beq
  \alpha_1 \not= \alpha_2 
  \label{eq:genericity}
\eeq
throughout the following consideration. 
We now formulate an elliptic analogue 
of the nonlinear Schr\"odinger hierarchy 
as the system of 
the Lax equations 
\beq
  [\rd_{t_n} - A_n(z),\; U(z)] = 0 
  \label{eq:Lax-tn-U}
\eeq
for the generating function $U(z)$ 
and the differential equations 
\beq
  \rd_{t_n}\gamma_s + \Tr \bfbeta_{n,s}\tp{\bfalpha_s} = 0, 
  \label{eq:rd-tn-gamma} \\
  \rd_{t_n}\tp{\bfalpha_n} + \tp{\bfalpha_s}A_n^{(s,1)} 
  = \kappa_{n,s}\bfalpha_s 
  \label{eq:rd-tn-alpha}
\eeq
for the Tyurin parameters.  Here $A_n^{(s,1)}$ denotes 
the constant term of the Laurent expansion of $A_n(z)$ 
at $z = \gamma_s$, i.e., 
\beqnn
  A_n^{(s,1)} 
  = \lim_{z\to\gamma_s}\left(
      A_n(z) - \frac{\bfbeta_{n,s}\tp{\bfalpha_s}}{z - \gamma_s}
    \right), 
\eeqnn
and $\kappa_{n,s}$ is a constant determined by 
the differential equation itself.  As in the case of 
(\ref{eq:rd-x-gamma}) and (\ref{eq:rd-x-alpha}), 
the two equations (\ref{eq:rd-tn-gamma}) and 
(\ref{eq:rd-tn-alpha}) for the Tyurin parameters 
are the necessary and sufficient conditions 
for the auxiliary linear system 
$\rd_{t_n}\psi(z) = A_n(z)\psi(z)$ to have 
a $2 \times 2$ matrix solution that is holomorphic 
at $z = \gamma_s$ and invertible except at these points.

\subsection{Zero-curvature equations} 

Commutativity of the time evolutions in $t_n$'s 
is by no means obvious from the construction.  
If one can derive the zero-curvature equations 
for $A_n(z)$'s, commutativity of the time evolutions 
is an immediate consequence.  As it turns out below, 
however, the zero-curvature equations in the present setting 
possess richer contents. 

Let us first derive a ``dual'' expression of 
the curvature components 
\beq
   F_{mn}(z) = [\rd_{t_m} - A_m(z),\; \rd_{t_n} - A_n(z)].  
\eeq
Let $A_n^{+}(z)$ denote the the ``tail'' part 
in the Laurent expansion of (\ref{eq:An-at-z=0}).  
Namely, 
\beq
  A_n^{+}(z) = A_n(z) - U(z)z^{-n}, 
\eeq
which has a Laurent expansion of the form 
\beqnn
  A_n^{+}(z) = (A_n^{(n+1)} - U_{n+1})z 
           + (A_n^{(n+2)} - U_{n+2})z^2 
           + \cdots, 
\eeqnn
where $A_n^{(m)}$'s denote the coefficients of 
the Laurent expansion 
\beqnn
  A_n(z) = \sum_{m=0}^\infty A_n^{(m)}z^{m-n} 
\eeqnn
of $A_n(z)$ at $z = 0$.  The Lax equations 
(\ref{eq:Lax-tn-U}) of $U(z)$ can be rewritten 
in the ``dual'' form 
\beq
  \rd_{t_n} U(z) 
  = [A_n^{+}(z) - U(z)z^{-n},\; U(z)] 
  = [A_n^{+}(z), U(z)]. 
  \label{eq:Lax-tn-U-dual}
\eeq
The curvature components turn out to have a similar 
dual expression as follows.  

\begin{lemma}
If the Lax equations (\ref{eq:Lax-tn-U}) are 
satisfied, the curvature components $F_{mn}(z)$ 
can be written in the dual form 
\beq
  F_{mn}(z) 
  = [\rd_{t_m} - A_m^{+}(z),\; \rd_{t_n} - A_n^{+}(z)]. 
  \label{eq:Fmn-dual}
\eeq
\end{lemma}

\proof 
Differentiating $A_m(z) = U(z)z^{-m} + A_m^{+}(z)$ by $t_n$ 
and using the Lax equation (\ref{eq:Lax-tn-U}), one has 
\beqnn
  \rd_{t_n}A_m(z) 
  &=& [A_n(z), U(z)]z^{-m} + \rd_{t_n}A_m^{+}(z) \nonumber \\
  &=& [A_n^{+}(z), U(z)]z^{-m} + \rd_{t_n}A_m^{+}(z), 
\eeqnn
and exchanging $m$ and $n$, 
\beqnn
  \rd_{t_m}A_n(z) 
  = [A_m^{+}(z), U(z)]z^{-n} + \rd_{t_m}A_n^{+}(z). 
\eeqnn
As for the commutator $[A_m(z),A_n(z)]$, 
\beqnn
  [A_m(z), A_n(z)] 
  &=& [U(z)z^{-m} + A_m^{+}(z),\; U(z)z^{-n} + A_n^{+}(z)] \nonumber \\
  &=& [A_m^{+}(z), U(z)]z^{-m} - [A_n^{+}(z), U(z)]z^{-n} 
      + [A_m^{+}(z), A_n^{+}(z)]. 
\eeqnn
Collecting these pieces yields the dual expression 
of the curvature.  
\qed

\begin{lemma}
If (\ref{eq:rd-tn-gamma}) and (\ref{eq:rd-tn-alpha}) 
are satisfied, the zero-curvature equations 
\beq
  [\rd_{t_m} - A_m(z),\; \rd_{t_n} - A_n(z)] = 0 
  \label{eq:zc-Am-An}
\eeq
can be derived from the Lax equations (\ref{eq:Lax-tn-U}). 
\end{lemma}

\proof
The following method of proof originates in 
the early work of Krichever and Novikov \cite{bib:KN80}. 
The curvature component $F_{mn}(z)$ is a matrix of 
meromorphic functions on $\Gamma$.  Suppose that 
$F_{mn}(z)$ turns out to satisfy the following conditions:  
\begin{enumerate}
\item $F_{mn}(z)$ is holomorphic at all points of $\Gamma$ 
other than possible poles at $\gamma_1,\gamma_2$.  
\item As $z \to 0$, $F_{mn}(z) = O(z)$. 
\item As $z \to \gamma_s$, $s = 1,2$, 
\beqnn
  F_{mn}(z) 
  = \frac{\bfbeta_{mn,s}\tp{\bfalpha_s}}{z - \gamma_s} 
    + O(1), 
\eeqnn
where $\bfbeta_{mn,s}$ is a two-dimensional column vector. 
\end{enumerate}
Such a matrix of function can be expressed as 
\beqnn
  F_{mn}(z) 
  = \sum_{s=1,2}\bfbeta_{mn,s}\tp{\bfalpha_s}
      (\zeta(z - \gamma_s) + \zeta(\gamma_s)).  
\eeqnn
By the residue theorem, the coefficients satisfy 
the relation 
\beqnn
  \sum_{s=1,2}\bfbeta_{mn,s}\tp{\bfalpha_s} = 0, 
\eeqnn
which, under the the genericity condition 
(\ref{eq:genericity}), imply that 
$\bfbeta_{mn,s} = 0$, hence $F_{mn}(z) = 0$.  
Thus the proof is reduced to confirming that 
$F_{mn}(z)$ does have the three properties.  
The first and second properties are now obvious; 
in particular, the dual expression of $F_{mn}(z)$ 
and the fact that $A_n^{+}(z) = O(z)$ imply that 
$F_{mn}(z) = O(z)$ as $z \to 0$.  What is left is 
to check the third property.  To this end, note that 
\beqnn
  F_{mn}(z) 
  = \left[ 
      \rd_{t_m} 
      - \frac{\bfbeta_{m,s}\tp{\bfalpha_s}}{z - \gamma_s} 
      - A_m^{(s,1)} + O(z - \gamma_s),\; 
      \rd_{t_n} 
      - \frac{\bfbeta_{n,s}\tp{\bfalpha_s}}{z - \gamma_s} 
      - A_n^{(s,1)} + O(z - \gamma_s) 
    \right] 
\eeqnn
as $z \to \gamma_s$.  Expanded to powers of $z - \gamma_s$, 
one can readily see, by (\ref{eq:rd-tn-gamma}), that 
the coefficient of $(z - \gamma_s)^{-2}$ vanishes.  
It is also easy to see, by (\ref{eq:rd-tn-alpha}), 
that the coefficient of $(z - \gamma_s)^{-1}$ is 
a rank-one matrix of the factorized form 
$\bfbeta_{mn,s}\tp{\bfalpha_s}$.  
\qed 

One can conversely derive the Lax equations 
(\ref{eq:Lax-tn-U}) from the zero-curvature equations. 

\begin{lemma}
The Lax equations (\ref{eq:Lax-tn-U}) can be derived 
from the zero-curvature equations (\ref{eq:zc-Am-An}).  
\end{lemma}

\proof 
Substituting $A_m(z) = A_m^{+}(z) + U(z)z^{-m}$ 
in the zero-curvature equation yields 
\beqnn
  [\rd_{t_m} - A_m^{+}(z) - U(z)z^{-m},\; 
   \rd_{t_n} - A_n(z)] = 0, 
\eeqnn
which one can further rewrite as 
\beqnn
  [\rd_{t_n} - A_n(z),\; U(z)] 
  = [\rd_{t_n} - A_n(z),\; 
     \rd_{t_m} - A_m^{+}(z)] z^m. 
\eeqnn
Since $A_m^{+}(z) = O(1)$ and $A_n(z) = O(z^{-n})$ 
as $z \to 0$, the right hand side of the last 
equation is $O(z^{m-n})$, so that 
\beqnn
  [\rd_{t_n} - A_n(z),\; U(z)] = O(z^{m-n}). 
\eeqnn
Letting $m \to \infty$, one obtains the Lax equation 
(\ref{eq:Lax-tn-U}) as expected.  \qed 

We thus eventually arrive at the following conclusion. 

\begin{theorem}
As far as (\ref{eq:rd-tn-gamma}) and (\ref{eq:rd-tn-alpha}) 
are satisfied, the Lax equations (\ref{eq:Lax-tn-U}) 
and the zero-curvature equations (\ref{eq:zc-Am-An}) 
are equivalent.  
\end{theorem}

Let us conclude the present consideration 
with a comment on (\ref{eq:rd-tn-gamma}) and 
(\ref{eq:rd-tn-alpha}).  Although these equations 
look somewhat distinct from the other equations, 
these equations themselves are directly related 
to the zero-curvature equations 
\beq
  [\rd_{t_n} - A_n(z),\; \rd_x - A(z)] = 0 
  \label{eq:zc-An-A}
\eeq
between $A_n(z)$ and $A(z)$.  Namely, if 
(\ref{eq:rd-x-gamma}) and (\ref{eq:rd-x-alpha}) 
are satisfied (this should be understood as  part of 
the definition of $A(z)$), (\ref{eq:rd-tn-gamma}) and 
(\ref{eq:rd-tn-alpha}) follow from these zero-curvature 
equations.  One can indeed derive these equations from 
the Laurent expansion of the left hand side of 
(\ref{eq:zc-An-A}) at $z = \gamma_s$.  In this respect, 
one may consider the zero-curvature equations 
(\ref{eq:zc-An-A}) as the defining equation of a hierarchy. 
This is indeed the way Krichever formulates a hierarchy.

\newsection{Riemann-Hilbert problem}

In this and next sections, we encounter various 
initial value problems with regard to the time variables 
$t = (t_1,t_2,\cdots)$, in which $t_1$ is identified 
with $x$.   For this reason, let us make the notations 
slightly more strict.  Namely, we write a $t$-dependent 
quantity always indicating its $t$-dependence explicitly 
as $A_n(t,z)$, $\gamma_s(t)$, $\alpha_s(t)$, etc.  
Otherwise, a quantity is understood to be independent 
of $t$.  

\subsection{Laurent series solution of auxiliary linear system} 

As a consequence of (\ref{eq:Fmn-dual}), we have the 
``dual'' zero-curvature equations 
\beq
  [\rd_{t_m} - A_m^{+}(t,z),\; \rd_{t_n} - A_n^{+}(t,z)] = 0. 
\eeq
These equations are the Frobenius integrability condition 
of the linear system 
\beq
  \rd_{t_n}\phi(t,z) = A_n^{+}(t,z)\phi(t,z). 
\eeq
One can redefine the generating function $\phi(t,z)$ 
to satisfy these equations as well.  

\begin{theorem}
Upon being suitably modified, the generating function 
$\phi(t,z)$ satisfies the forgoing linear system or, 
equivalently, 
\beq
  \rd_{t_n}\phi(z) = A_n(t,z)\phi(t,z) - \phi(t,z)Jz^{-n}. 
\eeq
In particular, 
\beq
  \psi(t,z) 
  = \phi(t,z)\exp\Bigl(\sum_{n=1}^\infty t_nJz^{-n}\Bigr) 
  \quad (t_1 = x) 
\eeq
gives a Laurent series solution of the auxiliary linear system 
\beq
  \rd_{t_n}\psi(t,z) = A_n(t,z)\psi(t,z). 
\eeq
\end{theorem}

\proof 
One can construct a Laurent series $\tilde{\phi}(t,z) 
= I + \tilde{\phi}_1 z + \cdots$ 
as a solution of the initial value problem 
\beqnn
  \rd_{t_n}\tilde{\phi}(t,z) = A_n^{+}(t,z)\tilde{\phi}(t,z), 
  \quad 
  \tilde{\phi}(t,z)|_{t_2=t_3=\cdots= 0} 
  = \phi(t,z)|_{t_2=t_3=\cdots=0}. 
\eeqnn
The Frobenius integrability condition of this system 
is ensured by the zero-curvature equation of $A_n^{+}(t,z)$'s. 
Moreover, since $A_n^{+}(t,z) = O(z)$, the solution 
persists to be of the form $I + O(z)$.  
Now consider the new Laurent series 
\beqnn
  \tilde{U}(t,z) = \tilde{\phi}(t,z)J\tilde{\phi}(t,z)^{-1}, 
\eeqnn
which satisfies the differential equations 
\beqnn 
  \rd_{t_n}\tilde{U}(t,z) 
  = [\rd_{t_n}\tilde{\phi}(t,z)\cdot\tilde{\phi}(t,z)^{-1},\; \tilde{U}(t,z)] 
  = [A_n^{+}(t,z), \tilde{U}(t,z)].  
\eeqnn
On the other hand, one knows that $U(t,z)$, too, 
satisfies differential equations of the same form, i.e., 
(\ref{eq:Lax-tn-U-dual}).  Since $\tilde{U}(t,z)$ and 
$U(t,z)$ have the same initial data at 
$t_2 = t_3 = \cdots = 0$, uniqueness of solution 
of the initial value problem implies that 
$\tilde{U}(t,z) = U(t,z)$, i.e., 
\beqnn
  U(t,z) = \tilde{\phi}(t,z)J\tilde{\phi}(t,z)^{-1}, 
\eeqnn
so that one can rewrite the foregoing differential 
equation for $\tilde{\phi}(t,z)$ as 
\beqnn
  \rd_{t_n}\tilde{\phi}(t,z) 
  = (A_n(t,z) - U(t,z)z^{-n})\tilde{\phi}(t,z) 
  = A_n(t,z) \tilde{\phi}(t,z) - \tilde{\phi}(t,z)Jz^{-n}. 
\eeqnn
Thus $\tilde{\phi}(t,z)$ turns out to fulfill 
all requirements. \qed

\subsection{Global solution of auxiliary linear system}

The Laurent series solution $\psi(t,z)$ of 
the auxiliary linear system, by its nature, 
carries no information on the global structure 
of $A_n(t,z)$'s on $\Gamma$.   To fill this gap, 
we now introduce another solution $\chi(t,z)$ 
that is globally defined on $\Gamma$ with 
several singular points.  As it turns out, 
these two distinct solutions of the same 
auxiliary linear system play the role of 
the Riemann-Hilbert (or factorization) pair 
in the usual nonlinear Schr\"odinger hierarchy.  

To avoid delicate problems, we assume in the following 
that the solutions of the hierarchy under consideration 
are (real or complex) analytic in a neighborhood of 
the initial point $t = 0$.  
 
$\chi(t,z)$, by definition, is a solution of 
the auxiliary linear system 
\beq
  \rd_{t_n}\chi(t,z) = A_n(t,z)\chi(t,z) 
\eeq
that satisfies the initial condition 
\beq
  \chi(0,z) = I. 
\eeq
Since the auxiliary linear system is a collection 
of ordinary differential equations, any solution remains 
nonsingular as far as the coefficients of the equations 
are nonsingular.  Consequently, if $z$ is in a subset 
of $\Gamma$ where $A_n(0,z)$'s are holomorphic, 
such a solution $\chi(t,z)$ does exists in a (possibly 
small) neighborhood of $t = 0$ in the $t$-space.  
Since all singularities of $A_n(0,z)$ on $\Gamma$ 
are located at the three points $0,\gamma_1(0),
\gamma_2(0)$, we can conclude that the singularities 
of $\chi(t,z)$ on $\Gamma$ are confined to a neighborhood 
of these three points as far as $t$ is sufficiently 
close to $0$.  

To elucidate the nature of singularities on $\Gamma$ 
more precisely, we expand $\chi(t,z)$ into a Taylor series 
at $t = 0$ and examine the Taylor coefficients as 
a function of $z$.  Note that this is reasonable, 
because this Taylor series has a nonzero radius of 
convergence as far as $z \not= 0,\gamma_1(0),\gamma_2(0)$.  

The Taylor coefficients of $\chi(t,z)$ at $t = 0$ 
can be evaluated by successively differentiating 
the differential equations as 
\beqnn
  \rd_{t_n}\chi(t,z) 
    &=& A_n(t,z)\chi(t,z), 
  \nonumber \\
  \rd_{t_m}\rd_{t_n}\chi(t,z) 
    &=& (\rd_{t_m}A_n(t,z) + A_n(t,z)A_m(t,z))\chi(t,z), 
  \nonumber \\
  \rd_{t_k}\rd_{t_m}\rd_{t_n} \chi(t,z) 
    &=& \Bigl(\rd_{t_k}\rd_{t_m}A_n(t,z) 
          + \rd_{t_k}(A_n(t,z)A_m(t,z)) 
    \nonumber \\
    && \mbox{} 
          + (\rd_{t_m}A_n(t,z))A_k(t,z) 
          + A_n(t,z)A_m(t,z)A_k(t,z)\Bigr)\chi(t,z), 
\eeqnn
etc.   Letting $t = 0$, we are left with 
a noncommutative polynomial of derivatives of $A_n$'s.  
We can deduce from these calculations the following 
precise information.  

\begin{lemma}
The derivatives 
$\rd_{t_{n_1}}\cdots\rd_{t_{n_p}}\chi(t,z)|_{t=0}$ 
of all orders of $\chi(t,z)$ at $t = 0$ are a matrix 
of meromorphic functions of $z$ on $\Gamma$ with poles 
at $z = 0,\gamma_1(0),\gamma_2(0)$ and holomorphic 
at other points.  As $z \to \gamma_s(0)$, $s = 1,2$, 
\beq
  \rd_{t_{n_1}}\cdots\rd_{t_{n_p}}\chi(t,z)|_{t=0} 
  = \frac{\bfbeta_{n_1,\cdots,n_1,s}(0)\tp{\bfalpha_s(0)}}
    {z - \gamma_s(0)} 
  + O(1), 
\eeq
where $\bfbeta_{n_1,\cdots,n_1,s}(0)$ is a two-dimensional 
constant column vector.  
\end{lemma}

\proof
As illustrated above, the derivatives of 
$\chi(t,z)$ of all order can be written as 
\beq
  \rd_{t_{n_1}}\cdots\rd_{t_{n_p}}\chi(t,z) 
  = A_{n_1,\cdots,n_p}(t,z)\chi(t,z). 
  \label{eq:rd-n1-np-chi}
\eeq
Differentiating this equation by $t_m$ yields 
the recurrence relations 
\beqnn
  A_{m,n_{1},\cdots,n_p}(t,z) 
  = \rd_{t_m}A_{n_1,\cdots,n_p}(t,z) 
    + A_{n_1,\cdots,n_p}(t,z)A_m(t,z). 
\eeqnn
for the coefficients $A_{n_1,\cdots,n_p}(t,z)$.  
One can prove, by induction on $p$, that 
$A_{n_1,\cdots,n_p}(t,z)$ is a matrix of 
meromorphic functions of $z$ on $\Gamma$ 
with poles at $z = 0,\gamma_1(t),\gamma_2(t)$, and 
\beq
  A_{n_1,\cdots,n_p}(t,z) 
  = \frac{\bfbeta_{n_1,\cdots,n_p,s}(t)\tp{\bfalpha_s(t)}}
         {z - \gamma_s(t)} 
    + O(1) 
  \label{eq:An1np-Laurent}
\eeq
as $z \to \gamma_s(t)$, where $\bfbeta_{n_1,\cdots,n_p,s}(t)$ is 
a two-dimensional column vector.  Assume that the Laurent expansion 
(\ref{eq:An1np-Laurent}) holds for $A_{n_1,\cdots,n_p}(t,z)$.  
The Laurent expansion of $A_{m,n_1,\cdots,n_p}(t,z)$ can be 
read off from the recurrence relation as 
\beqnn
\lefteqn{A_{m,n_1,\cdots,n_p}(t,z)} \nonumber \\
  &=& \bfbeta_{n_1,\cdots,n_p,s}(t)\tp{\bfalpha_s(t)} 
      \left(\rd_{t_m}\gamma_s(t) + \tp{\bfalpha_s(t)}\bfbeta_s(t)\right) 
      (z - \gamma_s(t))^{-2} \nonumber \\
  && \mbox{} 
     + \left(\rd_{t_m}\bfbeta_{n_1,\cdots,n_p,s}(t)\cdot\tp{\bfalpha_s(t)} 
         + \bfbeta_{n_1,\cdots,n_p,s}(t)\cdot\rd_{t_m}\tp{\bfalpha_s(t)} 
         + \mbox{}\right.  \nonumber \\
  && \phantom{\mbox{}+(} \left. 
         + A_{n_1,\cdots,n_p,s}^{(s,1)}\bfbeta_s(t)\tp{\bfalpha_s(t)} 
         + \bfbeta_{n_1,\cdots,n_p,s}(t)\tp{\bfalpha_s(t)}A_m^{(s,1)}(t) 
       \right)(z - \gamma_s(t))^{-1} \nonumber \\
  && \mbox{} 
     + O(1), 
\eeqnn
where $A_{n_1,\cdots,n_p}^{(s,1)}(t)$ denotes the constant 
term in the Laurent expansion (\ref{eq:An1np-Laurent}).  
By (\ref{eq:rd-tn-gamma}), the coefficient of 
$(z - \gamma_s(t))^{-2}$ vanishes; by (\ref{eq:rd-tn-alpha}), 
the terms containing $\rd_{t_m}\tp{\bfalpha_s(t)}$ and 
$\tp{\bfalpha_s(t)}A_m^{(s,1)}(t)$ in the coefficient 
of $(z - \gamma_s(t))^{-1}$ cancel out.  Thus 
$A_{m,n_1,\ldots,n_p}(t,z)$, too, turns out to have 
a Laurent expansion of the expected form.  This completes 
the proof of (\ref{eq:An1np-Laurent}).  Lastly, letting $t = 0$ 
in (\ref{eq:rd-n1-np-chi}), one eventually arrives at 
the statement of the lemma. 
 \qed 

All Taylor coefficients of $\chi(t,z)$ at $t = 0$ thus 
turn out to have poles at the same position, namely, 
the three points $0, \gamma_1(0), \gamma_2(0)$.  
Moreover, whereas the order of pole at $z = 0$ is 
unbounded, the poles at $z = \gamma_s(0)$, $s = 1,2$, 
are of the first order.  Accordingly, $\chi(t,z)$ 
has an essential singularity at $z = 0$ and 
simple poles at the other two points.  
The leading part of the Laurent expansion 
at $z = \gamma_s(0)$ takes the familiar form 
\beq
  \chi(t,z) 
  = \frac{\bfbeta_{\chi,s}(t)\tp{\bfalpha_s(0)}}
         {z - \gamma_s(0)} 
    + O(1), 
  \label{eq:chi-at-gamma}
\eeq
where $\bfbeta_{\chi,s}(t)$ is a two-dimensional 
column vector that depends on $t$.  Note that 
the pole of $\chi(t,z)$ at $z = \gamma_s(0)$ 
disappears when $t = 0$ (because $\chi(0,z) = I$).

Lastly, let us mention another important property 
of $\chi(t,z)$.  

\begin{lemma}
$\det\chi(t,z)$ is a meromorphic function on $\Gamma$ 
with simple poles at $z = \gamma_s(0)$, $s = 1,2$, 
and simple zeroes at $z =  \gamma_s(t)$, $s = 1,2$. 
$\tp{\bfalpha_s(t)}$ is a left null vector of 
$\chi(t,\gamma_s(t))$.  
\end{lemma}

\proof
The auxiliary linear system 
$\rd_{t_n}\chi(t,z) = A_n(t,z)\chi(t,z)$ 
induces the linear system 
\beqnn
  \rd_{t_n}\det\chi(t,z) = \Tr A_n(t,z) \det\chi(t,z) 
\eeqnn
for $n = 1,2,\ldots$.  Since 
$A_n(t,z) = U(t,z)z^{-n} + O(z)$ as $z \to 0$ and 
$\Tr U(t,z) = 0$, one finds that the coefficients 
of this linear system for $\det\chi(t,z)$ has 
no singularity at $z = 0$, but rather a zero, 
namely, 
\beqnn
  \Tr A_n(t,z) = O(z) \quad (z \to 0). 
\eeqnn
This implies that $\chi(t,z)$ has no singularity 
at $z = 0$.  In view of the initial condition 
$\chi(0,z) = I$, one can conclude that 
$\det\chi(t,z)|_{z=0} = 1$.  One can thus confirm 
that $\det\chi(t,z)$ is a meromorphic function 
on $\Gamma$ with poles at $z = \gamma_s(t)$, 
$s = 1,2$, and holomorphic at other points.  
Since the residue matrix of $\chi(t,z)$ at 
$z = \gamma_s(t)$ is a rank-one matrix, 
$\det\chi(t,z)$ has a simple pole there.  
The position of zeroes of $\det\chi(t,z)$ 
can be deduced from the the linear equation 
\beqnn
  \rd_x\chi(t,z) = A(t,z)\chi(t,z) 
\eeqnn
(or from any any member of the auxiliary linear 
system).  Extracting the residue at $z = \gamma_s(t)$ 
yields the relation 
\beqnn
  0 = \bfbeta_s(t)\tp{\bfalpha_s(t)}\chi(t,\gamma_s(t)) 
\eeqnn
which, because $\bfbeta_s(t) \not= \mathbf{0}$, 
reduces to the relation 
\beqnn
  \tp{\bfalpha_s(t)}\chi(t,\gamma_s(t)) = \mathbf{0}. 
\eeqnn
Thus $\tp{\bfalpha_s(t)}$ turns out to be 
a left null vector of $\chi(t,\gamma_s(t))$.  
On the other hand, rewriting the linear system as 
\beqnn
  A = \rd_x\chi(t,z)\cdot\chi(t,z)^{-1}, 
\eeqnn
one can see that the zeroes $\gamma_s(t)$ of 
$\det\chi(t,z)$ are simple.  If they are a multiple 
zero, the matrix $A$ will have a multiple pole; 
this contradicts the construction of the matrix $A$.  
\qed

These results show that $\chi(t,z)$ is exactly the solution 
mentioned in Krichever's lemma \cite[Lemma 5.2]{bib:Kr02}, 
namely a matrix solution holomorphic at the movable poles 
of $A(t,z)$.  

In summary, $\chi(t,z)$ has the following properties.  

\begin{theorem}\label{th:chi(t,z)}
$\chi(t,z)$ has an essential singularity at $z = 0$ 
and simple poles at $z = \gamma_s(0)$, $s = 1,2$, 
and is holomorphic at other points of $\Gamma$.  
As $z \to \gamma_s(0)$, $\chi(t,z)$ behaves as 
(\ref{eq:chi-at-gamma}) shows.  Moreover, 
$\det\chi(t,z)$ is a meromorphic function on 
$\Gamma$ with simple poles at $z = \gamma_s(0)$, 
$s = 1,2$, and simple zeros at $z = \gamma_s(t)$, 
$s = 1,2$.  $\tp{\bfalpha_s(t)}$ is a left null vector 
of $\chi(t,\gamma_s(t))$. 
\end{theorem}

\subsection{Riemann-Hilbert problem with degeneration points} 

We now have two distinct solutions of the same linear system, 
namely, the Laurent series solution $\psi(t,z)$ and 
the solution $\chi(t,z)$ carrying global information 
on $\Gamma$.  The ``matrix ratio'' of these these two solutions 
is a constant matrix, i.e.,  
\beqnn
  \rd_{t_n}\left(\chi(t,z)^{-1}\phi(t,z)
    \exp\Bigl(\sum_{n=1}^\infty t_nJz^{-n}\Bigr)\right) = 0. 
\eeqnn
Equating this matrix ratio with its value at $t = 0$, 
we are led to the relation 
\beq
  \chi(t,z)^{-1}\phi(t,z)\exp\Bigl(\sum_{n=1}^\infty t_nJz^{-n}\Bigr) 
  = \phi(0,z) 
\eeq
or, equivalently, 
\beq
  \phi(0,z) \exp\Bigl(- \sum_{n=1}^\infty t_nJz^{-n}\Bigr) 
  = \chi(t,z)^{-1}\phi(t,z).  
  \label{eq:RH} 
\eeq

The last relation may be thought of as a kind of 
Riemann-Hilbert problem concerning a small circle 
$|z| = a$ on the torus $\Gamma$.  The input of 
this problem are the initial values 
$\gamma_s(0),\alpha_s(0)$ and $\phi(0,z)$. 
The left hand side of (\ref{eq:RH}) is 
a $\mathrm{GL}(2,\CC)$-valued function on the 
circle, in other words, a $\mathrm{GL}(2,\CC)$ 
loop group element.  The problem is to factorize it 
to two factors.  The second factor $\phi(t,z)$ is 
a loop group element that can be extended to a matrix 
of holomorphic functions on the inside of the circle. 
The first factor $\chi(t,z)$ is a loop group element 
that can be similarly extended to the outside of 
the circle, but {\it not holomorphic everywhere}; 
$\chi(t,z)$ is required to have poles at 
$z = \gamma_s(0)$, $s = 1,2$, with the structure 
described in (\ref{eq:chi-at-gamma}).  Moreover,  
in addition to these poles, $\chi(t,z)$ have 
{\it degeneration points\/}, i.e., zeros of 
the determinant at $z = \gamma_s(t)$, $s = 1,2$.  
These zeroes are nothing but the poles of $A_n(t,z)$'s.  

Thus the Riemann-Hilbert problem relevant 
to the present setting is a Riemann-Hilbert problem 
with movable degeneration points (and extra fixed poles) 
on a torus.  A similar Riemann-Hilbert problem appears 
in Krichever's work \cite{bib:Kr78} on commutative rings 
of differential operators.  In that case, the Riemann-Hilbert 
problem is formulated on the ``spectral curve'' of 
the commutative ring under consideration, and the genus 
of the spectral curve can be an arbitrary positive integer.  

Krichever converts the Riemann-Hilbert problem to 
an integral equation and solves it by a standard procedure.  
The same method can be applied to the present setting as well, 
though we shall not seek this approach here.  An alternative 
approach, as demonstrated by Previato and Wilson \cite{bib:PW89}, 
is to translate the Riemann-Hilbert problem to the language 
of an infinite dimensional Grassmann variety.  We shall 
present this method in the next section.

\subsection{Back to hierarchy}

It will be instructive to show how to derive a solution 
of (\ref{eq:Lax-tn-U}), (\ref{eq:rd-tn-gamma}) and 
(\ref{eq:rd-tn-alpha}) from the Riemann-Hilbert problem.  
This is more or less parallel to the procedure that 
Krichever and Novikov employ in their work 
\cite{bib:KN78,bib:KN80}. 

Notice, first of all, that $\chi(t,z)$ is a matrix 
version of the ``vector Baker-Akhiezer function'' 
in their terminology.  This is an immediate consequence 
of the Riemann-Hilbert problem:  $\chi(t,z)$ has 
an essential singularity of the exponential type 
at $z = 0$, and fixed simple poles at $z = \gamma_s(0)$, 
$s = 1,2$.  Accordingly, the determinant $\det\chi(t,z)$ 
has zeros at $\gamma_s(t)$, $s = 1,2$,  that depend on $t$.  
Let us consider the generic situation where $\gamma_s(t)$'s 
are simple zeros of $\det\chi(t,z)$. The matrices $A_n(t,z)$, 
now defined by 
\beqnn
  A_n(t,z) = \rd_{t_n}\chi(t,z) \cdot \chi(t,z)^{-1}, 
\eeqnn
thereby has simple poles at $\gamma_s(t)$.  
As simple linear algebraic calculations show, 
the residue of $\chi(t,z)^{-1}$ at the degeneration point 
$\gamma_s(t)$ is a rank-one matrix.  Consequently, 
the residue of $A_n(t,z)$, too, is a rank-one matrix 
and takes the factorized form 
$\bfbeta_{n,s}(t)\tp{\bfalpha_s(t)}$ with 
a common vector $\bfalpha_s(t)$ independent of $n$.  
The dynamical Tyurin parameters $\gamma_s(t),\alpha_s(t)$, 
$s = 1,2$,  are thus obtained.  According to 
a general theorem of Krichever and Novikov 
(restated in Krichever's recent paper \cite{bib:Kr02}), 
these parameters satisfy the differential equations 
(\ref{eq:rd-tn-gamma}) and (\ref{eq:rd-tn-alpha}).  

One can now derive the Lax equation (\ref{eq:Lax-tn-U}) 
as follows.  Differentiating the Riemann-Hilbert 
relation (\ref{eq:RH}) yields another expression 
of $A_n(t,z)$, 
\beqnn
  A_n(t,z) = \rd_{t_n}\phi(t,z) \cdot \phi(t,z)^{-1} 
    + U(t,z)z^{-n}, 
\eeqnn
where $U(t,z)$ is defined as 
\beqnn
   U(t,z) = \phi(t,z)J\phi(t,z)^{-1}. 
\eeqnn
The Lax equations (\ref{eq:Lax-tn-U}) are thereby 
satisfied automatically.  Moreover, the second 
expression of $A_n(t,z)$ also shows the singular 
behavior of $A_n(t,z)$ as $z \to 0$: 
\beqnn
  A_n(t,z) = U(t,z)z^{-n} + O(z). 
\eeqnn
Thus $A_n(t,z)$'s turn out to have all properties 
that we have assumed in the construction of 
the hierarchy.

\newsection{Grassmannian perspective}

We here translate the Riemann-Hilbert problem to the language 
of an infinite dimensional Grassmann variety.  This leads 
to a mapping of the elliptic nonlinear Schr\"odinger 
hierarchy to a multi-time dynamical system on a subset 
(the set of dressed vacua) of the infinite dimensional 
Grassmann variety.

\subsection{Formulation of Grassmann variety}

Two different models of infinite dimensional 
Grassmann varieties have been used in the literature 
of integrable systems.  One is Sato's algebraic or 
complex analytic model based on a vector space of 
(formal or convergent) Laurent series \cite{bib:SS82}.  
The other is Segal and Wilson's functional analytic 
model based on the Hilbert space of square-integrable 
functions on a circle \cite{bib:SW85}.  Which to choose 
is rather a problem of taste; both of them work well 
in the present context.  Let us use Sato's model 
in the following.  Actually, Sato's formulation 
contains a continuous family of different models.  
Among them, we choose one of the presumably simplest models.  

Let $V$ denote the vector space of all $2 \times 2$ 
matrices of Laurent series 
\beqnn
  X(z) = \sum_{n=-\infty}^\infty X_nz^n, \quad 
  X_n \in \mathrm{gl}(2,\CC), 
\eeqnn
that converges in a neighborhood of $z = 0$ except at $z = 0$; 
$\mathrm{gl}(2,\CC)$ denotes the vector space of $2 \times 2$ 
complex matrices without any algebraic constraints.  
This vector space is a matrix analogue of 
$V^{\mathrm{ana}(\infty)}$ in Sato's list of models 
\cite{bib:SS82};  as noted therein, one can introduce 
a natural linear topology in this vector space.   

We construct an infinite dimensional Grassmann variety 
$\mathrm{Gr}$ from this vector space $V$ and 
the vector subspace 
\beq
  V_{+} = \{X(z) \in V \mid  \mbox{$X_n = 0$ for $n \le 0$}\} 
\eeq
of all $X(z) \in V$ that are holomorphic and vanish 
at $z = 0$.  The Grassmann variety $\mathrm{Gr}$ 
consists of all closed vector subspaces $W \subset V$ 
for which the composition of the inclusion map 
$W \hookrightarrow V$ and the canonical projection 
$V \to V/V_{+}$ is a Fredholm map of index $0$: 
\beq
  \mathrm{Gr} = \{W \subset V \mid 
    \dim\Ker(W \to V/V_{+}) = \dim\Coker(W \to V/V_{+}) < \infty\}.  
\eeq
The so called ``big cell'' $\mathrm{Gr}^\circ \subset \mathrm{Gr}$ 
is an open subset that consists of subspaces for which the map 
$W \to V/V_{+}$ is an isomorphism: 
\beq
  \mathrm{Gr}^\circ = \{W \in \mathrm{Gr} \mid W \simeq V/V_{+}\}. 
\eeq

\subsection{Vacuum and dressing} 

Following the idea of Previato and Wilson \cite{bib:PW89}, 
we now introduce a special element $W_0(\gamma,\alpha)$ 
of the big cell determined by constant Tyurin parameters 
$\gamma = (\gamma_1,\gamma_2)$ and $\alpha = (\alpha_1,\alpha_2)$.   
This is a matrix version of the ``vacuum'' that Previato 
and Wilson suggest to use for a holomorphic vector bundle 
in the Tyurin parametrization.  

\begin{lemma}\label{lem:W0-basis}
Let $\gamma = (\gamma_1,\gamma_2)$ be a pair of distinct 
points of $\Gamma$, $\gamma_1 \not= \gamma_2$, and 
$\alpha = (\alpha_1,\alpha_2)$ a pair of constants 
satisfying the genericity condition 
$\alpha_1 \not= \alpha_2$.  Then, for any integer 
$n \ge 0$ and the matrix indices $i,j = 1,2$, 
there is a unique $2 \times 2$ matrix $w_{n,ij}(z)$ 
of meromorphic functions on $\Gamma$ with 
the following properties: 
\begin{enumerate}
\item $w_{n,ij}(z)$ has poles at $z = 0,\gamma_1,\gamma_2$ 
and is holomorphic at other points.  
\item As $z \to 0$, $w_{n,ij}(z) = E_{ij}z^{-n} + O(z)$, 
where $E_{ij}$, $i,j = 1,2$, are the standard basis of 
$\mathrm{gl}(2,\CC)$.  
\item As $z \to \gamma_s$, $s = 1,2$, 
\beqnn
  w_{n,ij}(z) 
  = \frac{\bfbeta_{n,ij,s}\tp{\bfalpha_s}}
         {z - \gamma_s} 
    + O(1), 
\eeqnn
where $\bfalpha_s = \tp{(\alpha_s,1)}$, and 
$\bfbeta_{n,ij,s}$ is another two-dimensional 
constant column vector. 
\end{enumerate}
The subspace 
\beq
  W_0(\gamma,\alpha) 
  = \langle w_{n,ij}(z) \mid n \ge0,\; i,j = 1,2 \rangle 
\eeq
spanned by (the Laurent series of) $w_{n,ij}(z)$'s 
is an element of the big cell. 
\end{lemma}

\proof
One can confirm the existence and uniqueness of 
$w_{n,ij}(z)$ in the same way as the case of 
$A(z)$ and $A_n(z)$.  Since the leading terms 
$E_{ij}z^{-n}$ of the Laurent expansion at $z = 0$ 
are in one-to-one correspondence with the elements 
of the standard basis $\{E_{ij}z^{-n} \mid 
n \ge 0,\; i,j = 1,2\}$ of $V/V_{+}$, the linear map 
$W_0(\gamma,\alpha) \to V/V_{+}$ is obviously 
surjective.  To prove the injectivity, note that 
any element $X(z)$ of $W_0(\gamma,\alpha) \cap V_{+}$ 
is a matrix of functions holomorphic at all points 
of $\Gamma$ other than possible poles 
at $z = \gamma_1,\gamma_2$, behaves as 
\beqnn
  X(z) = \frac{\bfbeta_{X,s}\tp{\bfalpha_s}}{z - \gamma_s} 
         + O(1)  
  \quad 
\eeqnn
at these points (where $\bfbeta_{X,s}$ is 
a two-dimensional column vector), and has a zero 
at $z = 0$.  Such a matrix of function is equal 
to $0$ as one can see by the same reasoning 
as the proof of the zero-curvature equation 
(\ref{eq:zc-Am-An}).  Therefore 
$W_0(\gamma,\alpha) \cap V_{+} = \{0\}$, 
hence the injectivity of the linear map 
$W_0(\gamma,\alpha) \to V/V_{+}$ follows.  
\qed 

This special base point $W_0(\gamma,\alpha)$ of 
the big cell plays the role of vacuum in 
the ``dressing method.'' This is a complicated vacuum 
with nontrivial structure that stems from an underlying 
holomorphic vector bundle over $\Gamma$.   
We ``dress'' this vacuum to obtain an element $W$ 
of the big cell that represents a general solution 
of our hierarchy.  Dressing is achieved by multiplying 
a Laurent series $\phi(z)$ from the right side as 
\beq
  W = W_0(\gamma,\alpha)\phi(z), \quad 
  \phi(z) = I + \sum_{n=1}^\infty \phi_nz^n, \quad 
  \phi_n \in \mathrm{gl}(2,\CC).  
\eeq
Our goal in the following is to show that our hierarchy 
can be mapped to a multi-time dynamical system on the set 
\beq
\lefteqn{ \mathcal{M} = \{ W \in \mathrm{Gr}^\circ \mid  
      W = W_0(\gamma,\alpha)\phi(z),\;} \nonumber \\
  &&  \gamma = (\gamma_1,\gamma_2) \in \Gamma^2, \; 
      \alpha = (\alpha_1,\alpha_2) \in \CC^2,\; 
      \gamma_1\not= \gamma_2\,\; \alpha_1 \not= \alpha_2,\; 
      \phi_n \in \mathrm{gl}(2,\CC) \} 
\eeq
of these dressed vacua.

\subsection{Interpretation of Riemann-Hilbert problem}

We now translate the Riemann-Hilbert problem 
(\ref{eq:RH}) to the language of dressed vacua.  
Because of several reasons, the following consideration 
is limited to a small neighborhood of $t = 0$.  
Firstly, this is to ensure that the conditions 
$\gamma_1(t) \not= \gamma_2(t)$ and 
$\alpha_1(t) \not= \alpha_2(t)$ are satisfied; this issue 
is related to boundaries of the Tyurin parametrization of 
holomorphic vector bundles.  Secondly, if $t$ gets large, 
the dressed vacuum $W(t) \in \mathcal{M}$ representing  
a solution of (\ref{eq:RH}) can hit the boundary of 
the big cell, so that more careful analysis is required.  

The first step is the following.  

\begin{lemma}\label{lem:W0chi-sub-W0}
$W_0(\gamma(t),\alpha(t))\chi(t,z) 
  \subseteq W_0(\alpha(0),\gamma(0)).$ 
\end{lemma}

\proof
Let $w_{n,ij}(t,z)$, 
$n \ge 0$, $i,j=1,2$, denote the elements of the basis 
of $W_0(\gamma(t),\alpha(t))$ defined in Lemma 
\ref{lem:W0-basis}.  $w_{n,ij}(t,z)$ has poles at 
$z = 0,\gamma_1(t),\gamma_2(t)$, and behaves as 
\beqnn
  w_{n,ij}(t,z) 
  = \frac{\bfbeta_{n,ij,s}(t)\tp{\bfalpha_s(t)}}{z - \gamma_s(t)} 
    + O(1) 
\eeqnn
as $z \to \gamma_s(t)$.  Upon multiplication 
with $\chi(t,z)$, the poles at $z = \gamma_s(t)$ 
are cancelled out because $\tp{\bfalpha_s(t)}$ 
is a left null vector of $\chi(t,\gamma_s(t))$ 
(see Theorem \ref{th:chi(t,z)}).  
Thus one finds that $w_{n,ij}(t,z)\chi(t,z)$ has 
an essential singularity at $z = 0$, simple poles 
at $z = \gamma_s(0)$, $s = 1,2$, and is holomorphic 
at other points of $\Gamma$.   The leading part 
of the Laurent expansion at $z = \gamma_s(0)$ takes 
the form 
\beqnn
  w_{n,ij}(t,z)\chi(t,z) 
  = \frac{w_{n,ij}(t,\gamma_s(0))\bfbeta_{\chi,s}(t)
      \tp{\bfalpha_s(0)}}{z - \gamma_s(0)} 
    + O(1), 
\eeqnn
so that the residue matrix has such a factorized form 
as $(\mbox{column vector})\cdot\tp{\bfalpha_s(0)}$.  
One can thus confirm that $w_{n,ij}(t,z)\chi(t,z)$ 
fulfills all conditions to be an element of 
$W_0(\gamma(0),\alpha(0))$.  
\qed 

The next step is to show that the inclusion relation 
in this lemma is actually an equality.  To this end, 
we prove the following lemmas. 

\begin{lemma}
$\chi(t,z)^{-1}$ has an essential singularity at 
$z = 0$, simple poles at $\gamma_s(t)$, $s = 1,2$, and 
is holomorphic at other points.  As $z \to \gamma_s(t)$, 
\beq
  \chi(t,z)^{-1} 
  = \frac{\bfbeta_{\chi^{-1},s}(t)\tp{\bfalpha_s(t)}}
    {z - \gamma_s(t)} 
    + O(1), 
\eeq
where $\bfbeta_{\chi^{-1},s}(t)$ is a two-dimensional 
column vector.  
\end{lemma}

\proof
It is shown in Theorem \ref{th:chi(t,z)} that 
$\tp{\bfalpha_s(t)}$ is a left null vector of 
$\chi(t,\gamma_s(t))$.  A clue to the proof of 
the lemma is the fact that the left null space (i.e., 
the left zero-eigenspace) of $\chi(t,\gamma_s(t))$ 
is, actually, one-dimensional and spanned by 
$\tp{\bfalpha_s(t)}$.  If the left null space 
is two-dimensional, $\chi(t,\gamma_s(t))$ itself is 
a zero matrix, so that $\det\chi(t,z)$ has a double zero 
at $z = \gamma_s(t)$; this contradicts the present setting.  
[Remark: The same reasoning holds for an $r \times r$ 
analogue of the present case as well.  Namely, 
if the left null space of $\chi(t,\gamma_s(t))$ has 
$k$ dimensions, then $\det\chi(t,z)$ has a zero of 
the $k$-th order at $z = \gamma_s(t)$.] 
Bearing this fact in mind, one can prove the statement 
of the lemma as follows.  Theorem \ref{th:chi(t,z)}  
implies that $\gamma_s(t)$ is a simple zero of 
$\chi(t,z)^{-1}$.  Extracting the residue from 
the obvious identity $\chi(t,z)^{-1}\chi(t,z) = I$ 
yields the relation 
\beqnn
  \Res_{z=\gamma_s(t)}\chi(t,z)^{-1}dz \cdot 
  \chi(t,\gamma_s(t)) = 0. 
\eeqnn 
This implies that the residue matrix of $\chi(t,z)^{-1}$ 
at $z = \gamma_s(t)$ is a rank-one matrix of 
the factorized form $(\mbox{column vector})\cdot
(\mbox{row vector})$.  The row vector 
on the right side is accordingly a left null vector 
of $\chi(t,\gamma_s(t))$.  By the aforementioned fact, 
one can choose this row vector to be equal to 
$\tp{\bfalpha_s(t)}$.  Thus the residue matrix 
turns out to have a factorized form as shown 
in the statement of the lemma.  The other properties 
of $\chi(t,z)^{-1}$, too, can be readily derived from 
Theorem \ref{th:chi(t,z)}.  
\qed

\begin{lemma}
$\tp{\bfalpha_s(0)}$ is a left null vector of 
$\chi(t,z)^{-1}|_{z=\gamma_s(0)}$.  
\end{lemma}

\proof 
The identity $\chi(t,z)\chi(t,z)^{-1} = I$ 
yields the relation 
\beqnn
  \Res_{z=\gamma_s(0)}\chi(t,z)dz \cdot 
  \chi(t,z)^{-1}|_{z=\gamma_s(0)} = 0, 
\eeqnn
which, by (\ref{eq:chi-at-gamma}), takes the form 
\beqnn
  \bfbeta_{\chi,s}(t)\tp{\bfalpha_s(0)} 
  \chi(t,z)^{-1}|_{z=\gamma_s(0)} = 0.  
\eeqnn
Since $\bfbeta_{\chi,s}(t) \not= \mathbf{0}$, 
this implies that 
\beqnn
  \tp{\bfalpha_s(0)} \chi(t,z)^{-1}|_{z=\gamma_s(0)} = 0. 
\eeqnn
\qed

These lemmas show that the inverse matrix $\chi(t,z)^{-1}$ 
has essentially the same properties as $\chi(t,z)$ 
except that the position of poles and degeneration points 
are exchanged.   Consequently, one can repeat the proof of 
Lemma \ref{lem:W0chi-sub-W0}, replacing the role of 
$\chi(t,z)$, $W_0(\gamma(t),\alpha(t))$ and 
$W_0(\gamma(0),\alpha(0))$ with those of $\chi(t,z)^{-1}$, 
$W_0(\gamma(0),\alpha(0))$ and $W_0(\gamma(t),\alpha(t))$, 
to derive the inclusion relation 
\beqnn
  W_0(\gamma(0),\alpha(0))\chi(t,z)^{-1} 
  \subseteq W_0(\gamma(t),\alpha(t)). 
\eeqnn
Thus the equality 
\beq
  W_0(\gamma(t),\alpha(t))\chi(t,z) = W_0(\alpha(0),\gamma(0)). 
  \label{eq:W0chi-W0}
\eeq
follows as expected.  

Having this equality, one can readily convert 
the Riemann-Hilbert problem to the language of 
dressed vacua as follows.  The Riemann-Hilbert 
relation (\ref{eq:RH}) yields the relation 
\beqnn
  W_0(\gamma(t),\alpha(t))\phi(t,z) 
  = W_0(\gamma(t),\alpha(t))\chi(t,z)\phi(0,z) 
    \exp\Bigl(- \sum_{n=1}^\infty t_nJz^{-n}\Bigr). 
\eeqnn
By (\ref{eq:W0chi-W0}), $W_0(\gamma(t),\alpha(t))$ 
absorbs $\chi(t,z)$ to become $W_0(\gamma(0),\alpha(0))$.  
The outcome is the relation 
\beqnn
  W_0(\gamma(t),\alpha(t))\phi(t,z) 
  = W_0(\gamma(0),\alpha(0))\phi(0,z) 
    \exp\Bigl(- \sum_{n=1}^\infty t_nJz^{-n}\Bigr), 
\eeqnn
which means that the dressed vacuum 
$W(t) = W_0(\gamma(t),\alpha(t))\phi(t,z) \in \mathcal{M}$ 
obeys the exponential law 
\beq
  W(t) = W(0)\exp\Bigl(- \sum_{n=1}^\infty t_nJz^{-n}\Bigr). 
  \label{eq:exp-flow}
\eeq

Conversely, one can obtain a solution of 
the Riemann-Hilbert problem from the exponential flows 
(\ref{eq:exp-flow}) as follows.  (This is a variation of 
the dressing method of Previato and Wilson \cite{bib:PW89}.) 
Given a set of initial values $\gamma(0),\alpha(0)$ and 
$\phi(0,z)$,  let us consider the exponential flows 
(\ref{eq:exp-flow}) sending 
$W(0) = W_0(\gamma(0),\alpha(0)) \phi(0,z)$ to $W(t)$.  
A clue is, again, the fact that $W(t)$ remains 
in the big cell as far as $t$ is sufficiently small.  
In that case, the linear map $W(t) \to V/V_{+}$ is 
an isomorphism.  Let $\phi(t,z)$ denote the inverse image 
of $I \in V/V_{+}$ by this isomorphism.  
Being equal to $I$ modulo $V_{+}$, $\phi(t,z)$ is a Laurent 
series of the form 
\beqnn
  \phi(t,z) = I + \sum_{n=1}^\infty \phi_n(t)z^n. 
\eeqnn
On the other hand, as an element of 
\beqnn
  W(t) = W_0(\gamma(0),\alpha(0))\phi(0,z)
    \exp\Bigl(- \sum_{n=1}^\infty t_nJz^{-n}\Bigr), 
\eeqnn
$\phi(t,z)$ can also be written as 
\beqnn
  \phi(t,z) = \chi(t,z)\phi(0,z)
    \exp\Bigl(- \sum_{n=1}^\infty t_nJz^{-n}\Bigr) 
\eeqnn
with an element $\chi(t,z)$ of $W_0(\gamma(0),\alpha(0))$.  
Recalling the definition of $W_0(\gamma(0),\alpha(0))$, 
one finds that $\chi(t,z)$ is a matrix of functions with 
all properties in the statement of Theorem \ref{th:chi(t,z)}.  
The associated Tyurin parameters $(\gamma_s(t),\bfalpha_s(t))$ 
are determined as the position of zeros of $\chi(t,z)$ 
and the normalized left null vector of $\chi(t,z)$ at 
those degeneration points.  

Thus we have been able to show the following fundamental 
picture of our hierarchy as a dynamical system embedded 
in the Grassmann variety.  

\begin{theorem}
The elliptic analogue of the nonlinear Schr\"odinger hierarchy 
can be mapped, by the correspondence 
$W(t) = W_0(\gamma(t),\alpha(t))\phi(t,z)$, to a dynamical 
system on the set $\mathcal{M}$ of dressed vacua in 
the Grassmann variety $\mathrm{Gr}$.  The motion of $W(t)$ 
obeys the exponential law (\ref{eq:exp-flow}) .  
Conversely, the exponential flows on $\mathcal{M}$ yield 
a solution of the Riemann-Hilbert problem. (\ref{eq:RH}).  
\end{theorem}

Let us conclude this section with a few remarks.  
\begin{enumerate}
\item 
Our approach owes much to the work of Previato and Wilson 
\cite{bib:PW89}. They use a similar Grassmannian version 
of the dressing method as a tool  to reformulate the work 
of Krichever and Novikov \cite{bib:Kr78,bib:KN78,bib:KN80} 
on commutative rings of differential operators.  Accordingly, 
the detail of the dressing procedure is quite different 
from ours.  In particular, they take Krichever's 
``algebraic spectral data'' \cite{bib:Kr78} as the input; 
dressing is achieved by a matrix solution of linear 
differential equations determined by these data.  
In our case, the dressing matrix is the product of 
$\phi(0,z)$ and the exponential matrix generating 
the exponential flows (\ref{eq:exp-flow}).  
\item
In every aspect, the construction of the mapping to 
the Grassmann variety is related to the geometry of 
holomorphic vector bundles over $\Gamma$.  First 
of all, the Tyurin parameters $(\gamma_s(t),\alpha_s(t))$ 
themselves correspond to a holomorphic vector bundle 
that deforms as $t$ varies.  The subspace 
$W_0(\gamma,\alpha) \subset V$ can be identified 
with the space of holomorphic sections of 
the associated $\mathrm{sl}(2,\CC)$ bundle over 
the punctured torus $\Gamma \setminus \{z = 0\}$.  
$\phi(t,z)$ is related to changing local trivialization 
of this bundle at $z = 0$.  Note, in particular, 
that the data of local trivialization plays the role 
of dynamical variables.  This is to be contrasted 
with the work of Previato and Wilson; in their case, 
a set of functional parameters in the algebraic 
spectral data play a similar role in place of 
the data of local trivialization.  In this respect, 
our approach is more close to Li and Mulase's 
approach \cite{bib:Mu90,bib:LM97} to 
the classification of commutative rings of 
differential operators, in which the choice of 
local trivialization is treated as an independent data.  
\end{enumerate}

\newsection{Construction \`a la Enriquez and Rubtsov}

Enriquez and Rubtsov \cite{bib:ER99} parametrize 
the $\mathrm{sl}(2,\CC)$ Hitchin system on 
an algebraic curve of genus $g \ge 2$ \cite{bib:Hi90} 
by $3g$ (rather than $2g$) pairs $(\gamma_s,\alpha_s)$, 
$s = 1,\ldots,3g$, of Tyurin parameters.  The roles of 
parameters are also different from Krichever's formulation.  
Namely, whereas the directional vectors $\bfalpha_s 
= \tp{(\alpha_s,1)}$ remain dynamical, the poles $\gamma_s$ 
are fixed.  

We borrow their idea to construct another elliptic analogue 
of the nonlinear Schr\"odinger hierarchy.  This hierarchy 
has three pairs $(\gamma_s,\alpha_s)$, $s = 1,2,3$, 
as Tyurin parameters; $\gamma_s$ are constant and $\alpha_s$ 
are variables.  In addition to these Tyurin parameters, 
the hierarchy contains the nonlinear Schr\"odinger fields $u,v$.  
The foregoing consideration on the elliptic analogue of 
the Krichever type can be extended to this case with 
minimal modifications.

\subsection{Construction of $A$-matrix}

The $A$-matrix $A(z)$ is a $2 \times 2$ matrix of 
meromorphic functions on $\Gamma$ characterized 
by the following properties: 
\begin{enumerate}
\item $A(z)$ has poles at $z = 0,\gamma_1,\gamma_2,\gamma_3$ 
and holomorphic at other points.  
\item As $z \to 0$, 
\beqnn
   A(z) = \left(\begin{array}{cc} 
          z^{-1} & u \\
          v & - z^{-1} 
          \end{array}\right) 
        + O(z). 
\eeqnn
\item As $z \to \gamma_s$, $s = 1,2,3$, 
\beq
  A(z) 
  = \lambda_s \left(\begin{array}{cc}
      \alpha_s     & 1 \\
      - \alpha_s^2 & - \alpha_s 
    \end{array}\right) (z - \gamma_s)^{-1} 
  + O(1), 
\eeq
where $\lambda_s$ is a constant to be determined below. 
\end{enumerate}
One can write $A(z)$ itself more explicitly as 
\beq
  A(z) 
  = \sum_{s=1,2,3}
    \lambda_s \left(\begin{array}{cc}
      \alpha_s     & 1 \\
      - \alpha_s^2 & - \alpha_s 
    \end{array}\right) 
    (\zeta(z - \gamma_s) + \zeta(\gamma_s)) 
  + \left(\begin{array}{cc} 
      \zeta(z) & u \\
      v        & - \zeta(z) 
    \end{array}\right).  
\eeq
By the residue theorem, the coefficients 
have to satisfy the linear equations 
\beq
  \sum_{s=1,2,3}
    \lambda_s \left(\begin{array}{cc}
      \alpha_s     & 1 \\
      - \alpha_s^2 & - \alpha_s 
    \end{array}\right) 
  + \left(\begin{array}{cc} 
      1 & 0 \\
      0 & - 1  
    \end{array}\right) 
  = \left(\begin{array}{cc} 
      0 & 0 \\
      0 & 0 
    \end{array}\right).  
\eeq
This yields the three linear equations 
\beq
  \sum_{s=1,2,3}\lambda_s = 0, \quad 
  \sum_{s=1,2,3}\alpha_s\lambda_s = - 1, \quad 
  \sum_{s=1,2,3}\alpha_s^2\lambda_s = 0, 
\eeq
which can be solve for $\lambda_s$'s as 
\beq
  \lambda_1 = \frac{\alpha_3^2 - \alpha_2^2}{\Delta}, \quad 
  \lambda_2 = \frac{\alpha_1^2 - \alpha_3^2}{\Delta}, \quad 
  \lambda_3 = \frac{\alpha_2^2 - \alpha_1^2}{\Delta} 
\eeq
as far as the Vandermonde determinant 
\beqnn
  \Delta = \Delta(\alpha_1,\alpha_2,\alpha_3) 
  = \left|\begin{array}{ccc} 
      1 & 1 & 1 \\
      \alpha_1 & \alpha_2 & \alpha_3 \\
      \alpha_1^2 & \alpha_2^2 & \alpha_3^3 
    \end{array}\right| 
\eeqnn
does not vanish.   This condition 
\beq
  \Delta \not= 0 
\eeq
is the ``genericity condition'' in the present setting. 
We assume this condition throughout the consideration 
in the following.  

The Tyurin parameters are requirered to satisfy 
the differential equations (\ref{eq:rd-x-gamma}) and 
(\ref{eq:rd-x-alpha}).  Note that the residue matrices 
of $A(z)$ at $z = \gamma_s$ can be factorized as 
\beqnn
  \lambda_s 
  \left(\begin{array}{cc}
      \alpha_s     & 1 \\
      - \alpha_s^2 & - \alpha_s 
  \end{array}\right)
  = \lambda_s 
    \left(\begin{array}{c} 
      1 \\
      - \alpha_s 
    \end{array}\right)
    \left(\begin{array}{cc} 
      \alpha_s & 1 
    \end{array}\right), 
\eeqnn
so that the role of $\bfbeta_s$ is now played by 
$\lambda_s\tp{(1, - \alpha_s)}$. (\ref{eq:rd-x-gamma}) 
reduces to 
\beq
  \rd_x \gamma_s 
  = - \Tr \lambda_s 
      \left(\begin{array}{cc} 
        \alpha_s     & 1 \\ 
        - \alpha_s^2 & - \alpha_s 
      \end{array}\right) 
  = 0, 
\eeq
thus being consistent with the assumption that 
$\gamma_s$ are understood to be constant.  
On the other hand, (\ref{eq:rd-x-alpha}) takes 
the form 
\beq
  \rd_x \alpha_s 
  = \sum_{r \not= s} \lambda_r (\alpha_s - \alpha_r)^2 
     (\zeta(\gamma_s - \gamma_r) + \zeta(\gamma_r)) 
    + \alpha_s^2 u - 2\alpha_s\zeta(\gamma_s) + v 
\eeq
with the constant $\kappa_s$ uniquely determined as 
\beq
  \kappa_s 
  = \sum_{r \not= s} \lambda_r (\alpha_s - \alpha_r) 
     (\zeta(\gamma_s - \gamma_r) + \zeta(\gamma_r)) 
    + \alpha_s u - \zeta(\gamma_s). 
\eeq

\subsection{Construction of hierarchy} 

The construction of time evolutions is fully parallel 
to the previous case.   

Firstly, we construct a $2 \times 2$ matrix of 
generating functions 
\beqnn
  U(z) = \sum_{n=1}^\infty U_nz^n, \quad U_0 = J, 
\eeqnn
as a solution of the equations 
\beqnn
  [\rd_x - A(z),\; U(z)] = 0, \quad 
  U(z)^2 = I. 
\eeqnn
The coefficients $U_n$ are uniquely determined by 
a set of recurrence relations; the matrix elements 
thus turn out to be a differential polynomial of 
$\alpha_s$ ($s = 1,2,3$), $u$ and $v$.  

Having this generating function as local data 
at $z = 0$, we now proceed to the construction 
of the generators $A_n(z)$ of time evolutions.  
$A_n(z)$ is a $2 \times 2$ matrix of meromorphic 
functions on $\Gamma$ with the following properties: 
\begin{enumerate}
\item $A_n(z)$ has poles at $z = 0,\gamma_1,\gamma_2,\gamma_3$ 
and holomorphic at other points.  
\item A $z \to 0$, 
\beqnn
  A_n(z) = U(z)z^{-n} + O(z).  
\eeqnn
\item As $z \to \gamma_s$, $s = 1,2,3$, 
\beq
  A_n(z) 
  = \lambda_{n,s} 
    \left(\begin{array}{cc} 
      \alpha_s     & 1 \\
      - \alpha_s^2 & - \alpha_s 
    \end{array}\right) (z - \gamma_s)^{-1} 
  + O(1), 
\eeq
where $\lambda_{n,s}$ is a constant to be determined below. 
\end{enumerate}
$A_n(z)$ is uniquely determined by these conditions, 
and can be written as 
\beq
  A_n(z) 
  &=& \sum_{s=1,2,3} \lambda_{n,s} 
      \left(\begin{array}{cc} 
        \alpha_s     & 1 \\
        - \alpha_s^2 & - \alpha_s 
      \end{array}\right) 
      (\zeta(z - \gamma_s) + \zeta(\gamma_s)) \nonumber \\
  && \mbox{} 
    + \sum_{m=0}^{n-1}\frac{(-1)^{m}}{m!}\rd_z^m\zeta(z)U_{n-1-m} 
    + U_n.  
\eeq
The coefficients $\lambda_{n,s}$ are determined by 
the linear equations 
\beq
  \sum_{s=1,2,3} 
      \lambda_{n,s} 
      \left(\begin{array}{cc} 
        \alpha_s     & 1 \\
        - \alpha_s^2 & - \alpha_s 
      \end{array}\right) 
  + U_{n-1} = 0 
\eeq
or, equivalently, 
\beq
  \sum_{s=1,2,3}\lambda_{n,s} = - (U_{n-1})_{12}, 
  \nonumber \\
  \sum_{s=1,2,3}\alpha_s\lambda_{n,s} = - (U_{n-1})_{11} 
    = (U_{n-1})_{22}, 
  \nonumber \\
  \sum_{s=1,2,3}\alpha_s^2\lambda_{n,s} = - (U_{n-1})_{21}. 
\eeq
Of course, these linear equations are uniquely solvable 
as far as the genericity condition $\Delta \not= 0$ 
is satisfied.  

Lastly, the hierarchy is defined by the system of 
Lax equations 
\beq
  [\rd_{t_n} - A_n(z),\; U(z)] = 0 
\eeq
for $U(z)$ and the differential equations 
\beq
  \rd_{t_n}\tp{\bfalpha_s} + \tp{\bfalpha_s}A_n^{(s,1)} 
  = \kappa_{n,s}\tp{\bfalpha_s}, 
\eeq
for $\tp{\bfalpha_s} = (\alpha_s, 1)$.  
The differential equations for $\gamma_s$ reduce to 
\beq
  \rd_{t_n}\gamma_s 
  = - \Tr \lambda_{n,s} 
      \left(\begin{array}{cc} 
        \alpha_s     & 1 \\ 
        - \alpha_s^2 & - \alpha_s 
      \end{array}\right) 
  = 0 
\eeq
as expected.   One can derive the zero-curvature equations 
$[\rd_{t_m} - A_m(z),\; \rd_{t_n} - A_n(z)] = 0$ 
by the same procedure as in the previous case.

\subsection{Riemann-Hilbert problem and Grassmann variety}

The Riemann-Hilbert problem and the mapping to 
an infinite dimensional Grassmann variety can be 
derived in almost the same form as the previous case.  
The present case is conceptually rather simpler, 
because the poles $\gamma_s$ do not move.  
To avoid confusion, we again move to the convention 
that the $t$-dependence is always explicitly indicated 
as $A(t,z)$, $A_n(t,z)$, $\alpha_s(t)$, etc.  Note that
$\gamma_s$'s are constant throughout the present setting. 

The Riemann-Hilbert pair consists of a Laurent series 
solution $\psi(t,z)$ and a global solution $\chi(t,z)$ 
of the same auxiliary linear system.  The former takes 
the form 
\beqnn
  \psi(t,z) = \phi(t,z)\exp\Bigl(\sum_{n=1}^\infty t_nJz^{-n}\Bigr), 
  \quad 
  \phi(t,z) = I + \sum_{n=1}^\infty \phi_n(t)z^n. 
\eeqnn
The prefactor $\phi(t,z)$ is connected with the generating 
function $U(t,z)$ (the $t$-dependence is now shown explicitly) 
as $U(t,z) = \phi(t,z)J\phi(t,z)^{-1}$.  The second solution 
$\chi(t,z)$ of the auxiliary linear system is characterized 
by the initial condition $\chi(0,z) = I$.  One can prove, 
by the same technique as the previous case, that $\chi(t,z)$ 
has essential singularity at $z = 0$ and poles at 
$z = \gamma_1,\gamma_2,\gamma_3$, and behave as 
\beq
  \chi(t,z) 
  = \lambda_{\chi,s} 
    \left(\begin{array}{cc} 
      \alpha_s(0) & 1 \\
      - \alpha_s(0)^2 & - \alpha_s(0) 
    \end{array}\right)
    (z - \gamma_s)^{-1} 
  + O(1) 
\eeq
as $z \to \gamma_s$.  These two solutions 
$\chi(t,z),\psi(t,z)$ of the auxiliary linear system 
obeys a relation of the same form as (\ref{eq:RH}).  
On the other hand, since $A_n(z)$'s are trace-free, 
both $\chi(t,z)$ and $\phi(t,z)$ are now unimodular, 
i.e., 
\beq
  \det\chi(t,z) = \det\phi(t,z) = 1. 
\eeq
Consequently, unlike the previous case, $\chi(t,z)$ 
has no degeneration point. 

We use the same Grassmann variety $\mathrm{Gr}$ to 
embed the hierarchy.  The definition of the base point 
$W_0(\gamma,\alpha)$ for the present case, too, 
is essentially the same, except that we now use 
the three pairs $(\gamma_s,\alpha_s)$, $s = 1,2,3$, 
as the input.  The basis 
$\{w_{n,ij}(z) \mid n \ge 0,\; i,j=1,2 \}$ 
of $W_0(\gamma,\alpha)$ consists of the matrices 
$w_{n,ij}(z)$ of meromorphic functions on $\Gamma$ 
uniquely determined by these parameters as in 
the statement of Lemma \ref{lem:W0-basis};  
the third condition therein has to be modified as 
\beq
  w_{n,ij}(z) 
  = \lambda_{n,ij,s} 
    \left(\begin{array}{cc} 
      \alpha_s & 1 \\
      - \alpha_s^2 & - \alpha_s 
    \end{array}\right)
    (z - \gamma_s)^{-1} 
  + O(1) 
  \quad (z \to \gamma_s).  
\eeq
By the correspondence $\phi(t,z) \mapsto 
W(t) = W_0(\gamma,\alpha(t))\phi(t,z)$, the hierarchy  
is converted to a dynamical system on the set $\mathcal{M}$ 
of dressed vacua.  The motion of $W(t)$ again turns out 
to obey the same exponential law as (\ref{eq:exp-flow}).

\newsection{Conclusion}

We have elucidated the status of the two elliptic 
analogues of the nonlinear Schr\"odinger hierarchy 
in the Grassmannian perspective of Sato \cite{bib:SS82} 
and Segal and Wilson \cite{bib:SW85}.  Each of these systems 
are mapped to a dynamical system in the Grassmann variety 
$\mathrm{Gr}$.  The phase space of the dynamical system 
is the set $\mathcal{M}$ of dressed vacua 
$W = W_0(\gamma,\alpha)\phi(z)$.  The motion of 
the dressed vacuum $W(t) = W_0(\gamma(t),\alpha(t))\phi(t,z)$ 
under time evolutions of the hierarchy obeys a simple 
exponential law.   This is just the restriction of 
universal exponential flows on the Grassmann variety itself.  
Thus the situation is fully parallel to many classical 
soliton equations that have been understood in the 
Grassmannian perspective.  

It is straightforward to generalize the $2 \times 2$ system 
of the Krichever type to an $r \times r$ system \cite{bib:Kr02}.  
In that case, one has to use several $U$-matrices rather than 
a single one.  The Tyurin parameters consist of $r$ pairs 
$(\gamma_s,\bfalpha_s) \in \Gamma \times \PP^{r-1}$, 
$s = 1,\ldots,r$, of a point $\gamma_s$ of  $\Gamma$ 
and an $r$ dimensional directional vector $\bfalpha_s$.   
As a special case, one can obtain an elliptic analogue 
of the so called $N$ wave system, etc. 

If one does not insist on an explicit description of 
the system, one can generalize the results of this paper  
to an algebraic curve $\Gamma$ of genus $g$ with 
a marked point $P_0$.  The Tyurin parameters for 
the construction of the Krichever type consist of 
$2g$ pairs $(\gamma_s,\bfalpha_s) \in \Gamma \times \PP^1$ 
of a point of $\Gamma$ and a two dimensional directional vector 
\cite{bib:Kr02}.  The construction \`a la Enriquez and Rubtsov 
requires $3g$, rather than $2g$, pairs of Tyurin parameters 
\cite{bib:ER99}.  Upon choosing a local parameter $z$ 
in a neighborhood of $P_0$, one can start the construction 
of the fundamental matrix $A(P)$ ($P \in \Gamma$) of 
meromorphic functions on $\Gamma$ and the matrix $U(z)$ 
of Laurent series.  A convenient choice of $z$ is 
to define it as the (multivalued) primitive function 
$z(P) = \int_{P_0}^P \omega$ of a holormophic differential 
$\omega$ on $\Gamma$ without zero at $P_0$. 
The matrices $A_n(P)$, like $A(P)$, are characterized 
by a set of conditions on the poles.  Namely, 
they are matrices of meromorphic functions on $\Gamma$ 
with poles at $P_0$ and $\gamma_s$'s, and behave as 
\beqnn
  A_n(P) = U(z(P))z(P)^{-n} + O(z(P)) \quad (P \to P_0), 
  \nonumber \\
  A_n(P) = \frac{\bfbeta_{n,s}\tp{\bfalpha_s}}{z(P) - z(\gamma_s)} 
           + O(1) \quad (P \to \gamma_s). 
\eeqnn
The existence and the uniqueness of these matrices 
are ensured by the Riemann-Roch theorem.  
Formulating these systems in a more explicit 
form is a problem left for future research.  

Lastly, let us mention some other approaches to 
soliton equations associated with algebraic curves.  
Ben-Zvi and Frenkel \cite{bib:BF01} and 
Levin, Olshanetsky and Zotov \cite{bib:LOZ03} 
propose to construct those equations as a $1 + 1$ 
dimensional analogue of the Hitchin systems \cite{bib:Hi90}.  
The framework of Ben-Zvi and Frenkel is conceptually 
similar to ours, though they use a Grassmann variety 
in a different way.  The work of Li and Mulase 
\cite{bib:Mu90,bib:LM97} is also closely related to 
the present issue.  Our construction of dressed vacua 
has obviously a counterpart in their description of 
commutative rings of differential operators in 
the language of infinite dimensional Grassmann variety.

\subsection*{Acknowledgements}
I would like to thank Takeshi Ikeda and Takashi Takebe 
for discussion and many comments. 
This work was partly supported by 
the Grant-in-Aid for Scientific Research 
(No. 14540172) from the Ministry of Education, 
Culture, Sports and Technology.

\end{document}